\documentclass[12pt]{article}\usepackage{natbib,amssymb,psfig}
\newcommand{\beq}{\begin{equation}}
\newcommand{\eeq}{\end{equation}}
\newcommand{\bea}{\begin{eqnarray}}
\newcommand{\eea}{\end{eqnarray}}
\newcommand{\mdot}{{\dot m}}
\newcommand{\dt}[1]{\frac{\partial{#1}}{\partial t}}
\newcommand{\dx}[1]{\frac{\partial{#1}}{\partial x}}
\newcommand{\rem}[1]{ }
\newcommand{\la}{\lesssim}
\newcommand{\ga}{\gtrsim}
\newcommand{\apj}{{\it ApJ}}
\newcommand{\apjs}{{\it ApJS}}
\newcommand{\aap}{{\it A\&A}}
\newcommand{\mnras}{{\it MNRAS}}
\newcommand{\prl}{{\it PRL}}
\newcommand{\araa}{{\it ARA\&A}}
\newcommand{\pasj}{{\it Proc. R. Soc. Japan}}
\bibpunct[,]{(}{)}{;}{a}{}{,}
\begin{document}

\title{Hot Radiative Accretion onto a Spinning Neutron Star}

\author{Mikhail V. Medvedev 
\thanks{
Also at the Institute for Nuclear Fusion, RRC ``Kurchatov
Institute'', Moscow 123182, Russia}
\\
Department of Physics and Astronomy, \\
University of Kansas, KS 66045}

\maketitle

\begin{abstract}
A new type of self-similar hot viscous radiative accretion flow onto 
a rapidly spinning neutron star has recently been discovered. This 
``hot brake'' 
flow forms in the two-temperature zone (close to a central object), 
but at a sufficiently low accretion rate and a high spin it may extend 
in the radial direction beyond $\sim300$ Schwarzchild radii into 
a one-temperature zone. When the spin of the star is small enough, 
the flow transforms smoothly to an advection-dominated accretion flow.

The properties of the hot brake flow are rather exceptional and 
surprising. All gas parameters (density, angular velocity, temperature, 
luminosity, angular momentum flux) except for the radial velocity 
are independent of the mass accretion rate; these quantities do 
depend sensitively on the spin of the neutron star.  The gas angular 
momentum is transported outward under most conditions, hence
the central star is nearly always spun-down.
The luminosity of the hot brake flow arises from the rotational energy
that is released as the star is braked by viscosity. The
contribution from gravity is small, therefore the radiative efficiency
may be arbitrarily large as $\dot M\to0$.  
We demonstrate that the flow is also convectively stable
and is unlikely to produce a strong outflow or wind.

The hot brake flow is cooling-dominated (via Bremsstrahlung) and, 
hence, might be thermally unstable. The analysis of thermal 
conduction in the hot gas shows that thermal transport is 
collisionless (non-Spitzer) and occurs via free streaming 
of electrons along tangled magnetic field lines.  We find that 
conduction is strong enough to make the flow thermally stable. 

The very fact that the density, temperature and angular velocity 
of the gas at any radius in the hot brake flow are completely 
independent of the outer and inner (except for the star spin) 
boundary conditions implies that the flow cannot be smoothly 
matched to a general external medium as well as to general 
conditions on the star surface. We demonstrate that there are 
two extra self-similar solutions: one bridges the gap between the
original solution and the external medium, and another 
represents a boundary layer between the bulk of the flow and 
the star surface, in which the gas temperature 
rapidly drops while the density builds up.

Finally, we briefly discuss that a hot brake flow may form 
around other rapidly spinning compact objects: white
dwarfs and black holes.
\end{abstract}

\section{Introduction}

Accretion flows around compact objects frequently radiate significant
levels of hard X-rays, indicating the presence of hot optically-thin
gas in these systems.  This has motivated the study of hot accretion
flows around compact stars.

\citet{ZS69} and \citet{AW73} considered spherically free-falling 
plasma impinging on the surface of a neutron star (NS).
They calculated the penetration depth of the falling protons and made
preliminary estimates of the radiated spectrum.  Their ideas were
followed up by a number of later authors, e.g., \citet{Tur94},
\citet{Zam95}, and \citet{Zan98}, who carried out more
detailed computations of spectra.

A fluid approach to spherical accretion onto a NS was pioneered by
\citet{SS75}, who worked out the structure of the
standing shock in a spherical flow and computed the two-temperature
structure of the post-shock gas and the resulting spectrum.  The
equivalent problem for an accreting white dwarf was analyzed by
\citet{KL82}.  In related work, \citet{CS97}
described the hydrodynamics of spherical accretion onto black holes
and NSs, but without including radiation processes.

All of the studies listed above involve inviscid spherical flows and
include shocks of some kind, caused by spherically accreting  matter
crashing on the surface of the accreting star.  However, if the
accretion flow has angular momentum and viscous transport, there are
strong reasons why a shock is not expected.  A shock introduces a
causal discontinuity between the accreting star and the inflowing gas.
While such a discontinuity is not a problem for a spherical flow,
\citet[ see also \citealp{PN92}]{P77} argued that it leads to a 
serious physical inconsistency if the flow is rotating. \citet{PN92}
showed that a causal viscosity prescription leads to
consistent viscous accretion solutions without shocks.  \citet{PS01} 
have calculated detailed boundary layer solutions for
accretion onto a neutron star, but these solutions are based on the
cool thin accretion disk model and are not relevant for the hot flows
we discuss here.

One situation in which a shock is possible with a rotating flow is if
the surface of the accreting star lies inside the marginally stable
orbit of a thin disk.  One then expects ``gap accretion'' with a shock
\citep{KW91}, and there is no inconsistency in having a
shock.  The marginally stable orbit appears not to play an important
role in hot quasi-spherical flows (see \citealp{N97} and
\citealp{C97}).  It is, therefore, not clear that
one would necessarily have a shock with a hot flow even if the
accreting star were very compact.

\citet{D01} modified the model of \citet{ZS69} 
by considering a rotating advection-dominated accretion
flow (see below) around a NS.  Their model represents an improvement
on the earlier work since it includes angular momentum and viscosity,
but it still invokes a shock of some kind, since there is a
discontinuity at the radius where the hot ADAF meets the surface of
the NS.

Following the seminal work on two-temperature accretion flows by
\citet[ the SLE solution]{SLE76}, other hot
solutions were discovered to describe accretion onto black holes: the
advection-dominated accretion flow (ADAF)
\citep{I77,R82,NY94,NY95a,NY95b,Abram+95}, the advection-dominated
inflow-outflow solution (ADIOS) \citep{BB99,NY94,NY95a}, and the
convection-dominated accretion flow (CDAF) \citep{NIA99,QG99}.  All of
these solutions describe rotating flows with viscosity and angular
momentum transfer.  The relevance of the solutions for accretion onto
a NS is, however, unclear.

Recently, \citet{MN01} discovered a rotating shock-free solution of
the viscous fluid equations that corresponds to hot quasi-spherical
accretion onto a rapidly spinning NS.  We refer to this solution as a
``{\em hot brake flow},'' since the gas is hot (the temperature is
nearly virial) and it spins the NS down (in previous works it was called
the ``hot settling flow'').
The solution could equally well be described as a ``{\em hot
atmosphere}'' since the solution is to first approximation static, and
accretion represents only a small perturbation on the static solution
(as is probably true for any settling flow).  To our knowledge, the
hot brake flow is the only solution for accretion onto a NS
presently available that does not involve a discontinuity near the
surface of the star.  The hot brake flow should not be confused with a
boundary layer which forms in the very vicinity of the stellar surface
(e.g., \citealp{NP93}) and which is characterized by a high gas
density and steep spatial gradients of physical parameters. The hot
brake flow forms {\em above} the boundary layer and extends radially to a
large distance, typically thousands of stellar radii or more.
Following up the work by \citep{DP81}, \citet{I01,I03} has recently 
presented a subsonic hot accretion flow around a magnetized neutron 
star in the propeller state. The main difference between the two cases 
is that in the hot brake flow, heating and cooling balance each other, 
whereas in the subsonic propeller the heating rate of the accreting gas
due to viscous dissipation is much larger than the radiative cooling
rate. Hence, the latter solution shows some resemblance to the
advection-dominated (or convection-dominated) flow.

The hot brake flow exists at rather low accretion rates, smaller
than a few percent of Eddington. The flow is subsonic everywhere
(which is why it does not form a shock near the NS surface).  Because
the accreting gas has a low density and high temperature, the particle
mean free path is larger than the local radius of the flow
and the gas is essentially collisionless.
Viscosity plays a very important role; indeed, the flow is powered by
the rotational energy of the central accretor which is braked by
viscous torques. A very interesting property of the flow is that,
except for the inflow velocity, all gas properties, such as density,
temperature, angular velocity, luminosity, and angular momentum flux,
are independent of the mass accretion rate, as might be expected from
the earlier comment that accretion behaves like a minor perturbation
on an intrinsically static solution; the flow properties do depend on
the NS spin (see \citealp{MN01} for more details). 

In this paper 
we present the analytical self-similar solution describing this
{\it hot brake flow}, study its stability and observable properties. 
We also present the solution that matches the hot break flow to an
arbitrary external medium and the boundary layer solution that
matches the flow to the star surface.
All our theoretical results are confirmed with numerical solutions
of the appropriate set of hydrodynamic equations.

\section{Basic considerations}
\subsection{The mathematical model}

We consider gas accreting viscously onto a compact spinning object
with a surface. The central object has a radius $R_*$, a mass
$M_*=m M_{\rm Sun}$, and an angular velocity $\Omega_*=s \Omega_K(R_*)$,
where $\Omega_K(R)=(GM_*/R^3)^{1/2}$ is the Keplerian angular velocity
at radius $R$.  We measure the accretion rate in Eddington units,
$\dot m=\dot M/\dot M_{\rm Edd}$, and the radius in Schwarzchild
units, $r=R/R_g$, where $\dot M_{\rm
Edd}=1.39\times10^{18}m$~g~s$^{-1}$ (corresponding to a radiative
efficiency of 10\%) and $R_g=2GM_*/c^2$.
We use the height-integrated form of the viscous hydrodynamic equations 
\citep{I77,A+88,Paczynski91,NY94}:
\bea
\displaystyle &\displaystyle \dot M=4\pi R^2\rho v , 
& \label{mdot}\\
&\displaystyle v\frac{d v}{d R}=\left(\Omega^2-\Omega_K^2\right)R
-\frac{1}{\rho}\frac{d}{dR}\left(\rho c_s^2\right) , 
& \label{mom}\\
&\displaystyle 4\pi\nu\rho R^4\frac{d\Omega}{dR} 
=\dot J-\dot M\Omega R^2 , 
& \label{omega}\\ 
&\displaystyle \rho v T_p\frac{d s_p}{dR}
=\frac{\rho vc^2}{(\gamma_p-1)}\frac{d\theta_p}{dR}
-vc^2\theta_p\frac{d\rho}{dR}=q^+-q_{\rm Coul} , 
& \label{energy-p}\\
&\displaystyle \rho_e v T_e\frac{d s_e}{dR}
=\frac{\rho_e vc^2}{(\gamma_e-1)}\frac{d\theta_e}{dR}
-vc^2\theta_e\frac{d\rho_e}{dR}=q_{\rm Coul}-q^- ,
& \label{energy-e}
\eea 
where $\rho$ is the mass density of the accreting gas, $v$ is the
radial infall velocity, $\Omega$ is the angular velocity,
$c_s^2=c^2(\theta_p+\theta_e m_e/m_p)$ is the square of the isothermal
sound speed, $T_{p,e}$ are the temperatures of protons and electrons,
$\theta_{p,e}=k_BT_{p,e}/m_{p,e}c^2$ are the corresponding
dimensionless temperatures, $\nu$ is the gas viscosity, 
$\dot J$ is the rate of accretion of angular momentum,
$s_p$ and $s_e$ are the specific entropies of the proton and electron
fluids, $\rho_e\simeq(m_e/m_p)\rho$ is the mass density of the
electron fluid, $\gamma_p$ and $\gamma_e$ are the adiabatic indexes of
protons and electrons (which, in general, may be functions of $T_p$
and $T_e$), and $q^+$, $q^-$, and $q_{\rm Coul}$ are the viscous
heating rate, radiative cooling rate, and energy transfer rate from
protons to electrons via Coulomb collisions, per unit mass. 
We have assumed that all viscous heat goes into the proton component.
Equations (\ref{mdot})--(\ref{energy-e}) describe the conservation of mass,
radial momentum, angular momentum, proton energy and electron energy,
respectively.

We employ the usual $\alpha$ prescription for the kinematic
coefficient of viscosity, which we write as 
\beq 
\nu=\alpha c_s H\approx \alpha c_s R.
\label{nu}
\eeq 
Often, in accretion problems, one makes use of the relation
$H=c_s/\Omega_K$ and writes $\nu=\alpha c_s^2/\Omega_K$.  This
prescription is equivalent to $\nu=\alpha c_s H\approx\alpha c_s R$ 
in the regime of the \citet{MN01}
hot brake flow.  However, in the outer regions of the flow, where
the other solutions described in the following sections appear, $H$ is 
much less than $c_s/\Omega_K$, and $\nu=\alpha c_s^2/\Omega_K$ is not 
a good approximation.  Equation (\ref{nu}) is a superior prescription and is
physically better motivated over a wide range of conditions (so long
as the flow is quasi-spherical).

We assumed that the flow is hot and quasi-spherical, which generally
requires a low mass accretion rate (see \citealp{NMQ97}). The
accreting gas has nearly the virial temperature, i.e., $c_s^2\sim GM_*/R
\sim (\Omega_K R)^2$, and the local vertical scale height
$H=c_s/\Omega_K$ is comparable to the local radius $R$.  We may then
use the height integrated hydrodynamic equations for a steady,
rotating, axisymmetric flow, and for simplicity we may set $H=R$ \citep{MN01}.

In the case of accretion onto a NS we expect the flow to slow down as
it settles on the stellar surface, and we expect the density in this
settling zone to be significantly higher than for a black hole (BH) 
accretion.  The increased
density would cause more efficient transfer of energy from protons to
electrons via Coulomb collisions and more efficient radiation from the
electrons.  As we show below, this leads to a flow in which $q^+$,
$q^-$ and $q_{\rm Coul}$ are all of the same order, which is very
different from the case of a BH ADAF, where $q^+\gg q^-,\, q_{\rm Coul}$. 
 Another feature of the settling zone, again the result of the
large density, is that optically thin bremsstrahlung cooling (which is
sensitive to $\rho$) dominates over self-absorbed synchrotron cooling.
We therefore neglect synchrotron emission in our analysis.  For
simplicity, we neglect also thermal conduction (we will include this effect 
in the discussion of thermal stability).  

The temperature of the gas determines the efficiency of the Coulomb 
energy transfer from the protons to the electrons and the rate of 
Bremsstrahlung cooling of the electrons. The balance between them
defines whether the gas is in the two temperature regime (when the 
temperatures are high and the Coulomb collisions are very rare) or
in the one-temperature regime (then temperatures are lower). 

The two-temperature regime occurs closer to the central object, where
the virial temperature high enough and the electrons become relativistic.
In the the inner region of the flow, $R_*\la R\la 100R_*$.
we expect a two-temperature plasma, with $T_p>T_e$, in which
the electrons are relativistic and the protons are non-relativistic:
$\theta_e\gg1,\ \theta_p\ll1$.  The viscous heating rate of the gas,
the energy transfer rate from the protons to the electrons via Coulomb
collisions, and the cooling rate of the electrons via bremsstrahlung
emission are given by
\bea 
q^+&=&\nu\rho R^2\left(\frac{d\Omega}{dR}\right)^2,
\label{q+}\\
q_{\rm Coul}&=&Q_{\rm Coul}\,\rho^2\frac{\theta_p}{\theta_e} ,\quad
Q_{\rm Coul}=4\pi r_e^2 \ln{\Lambda} \frac{m_ec^3}{m_p^2}, 
\label{qc}\\
q^-&=&Q_{\rm ff,R}\,\rho^2\theta_e ,\quad
Q_{\rm ff,R}=48\alpha_f r_e^2\frac{m_ec^3}{m_p^2} ,
\label{qff}
\eea 
where $\alpha_f$ is the fine structure constant, $r_e$ is the
classical electron radius, $\ln{\Lambda}\simeq20$ is the Coulomb
logarithm, $c_s^2\simeq c^2\theta_p$, and we have neglected logarithmic
corrections to the relativistic free-free emissivity. The subscript
``R'' in $Q_{\rm ff,R}$ denotes relativistic Bremsstrahlung.

For $R>100 R_*$, both protons and electrons are non-relativistic
and have nearly the same temperature $T_p-T_e\ll T_p,\,T_e$.  
The free-free cooling takes the form
\beq
q^-= Q_{\rm ff,NR}\rho^2\theta_e^{1/2}, \quad 
Q_{\rm ff,NR}=5\sqrt{2}\pi^{-3/2}\alpha_f\sigma_T\frac{m_ec^3}{m_p^2},
\eeq
where $\sigma_T$ is the Thompson cross-section, and the subscript
${\rm NR}$ stands for non-relativistic.  Since the gas is effectively
one-temperature, we have
\beq 
q^+\simeq q^- 
\eeq 
because the Coulomb transfer rate, $q_{\rm Coul}$,
is proportional to $(T_p-T_e)$ and
can adjust itself to have the right magnitude with small adjustments
of the two temperatures.  In the non-relativistic regime, $\theta_e<1,\ 
\theta_p\sim (m_e/m_p)\theta_e\ll1$, and the Coulomb transfer rate is
\beq
q_{\rm Coul}=\frac{3}{\sqrt{2\pi}}\frac{m_e}{m_p}\frac{\sigma_Tc}{m_p^2}
\ln\Lambda\,\rho^2\frac{kT_p-kT_e}{\theta_e^{3/2}e^{-1/\theta_e}}.
\eeq
From the condition $q_{\rm Coul}\simeq q^-$ it follows that
\beq
kT_p-kT_e\simeq\frac{10}{3\pi^2}\frac{\alpha_f}{\ln\Lambda}\sqrt{m_pm_e}c^2
\theta_e^2e^{-1/\theta_e},
\eeq
which is exponentially small for $\theta_e<1$.

The set of equations (\ref{mdot})--(\ref{energy-e}) must satisfy certain 
boundary conditions at the neutron star.  First, as the flow approaches the
surface of the star at $R=R_*$, the radial velocity must become
very much smaller than the local free-fall velocity.  Second, the
angular velocity must approach the angular velocity of the star $\Omega_*$.  

The radius of the star, and its spin, are the two principal boundary
conditions applied at the inner edge of the accretion flow.  We assume
that the star is unmagnetized, so there are no magnetospheric effects
to consider.  Two outer boundary conditions, namely the temperature
and angular velocity of the gas, are determined by the properties of
the gas as it is introduced into the accretion flow on the outside
(we will discuss this in subsequent sections). 
 An additional important boundary condition is the
mass accretion rate $\dot M$, which is determined by external
conditions and which we take to be constant.

\subsection{Numerical solution}

The system of equations given above with appropriate boundary conditions
has been solved using the relaxation method on a highly non-uniform
grid (in order to resolve the thin boundary layer, where the density
rises by few orders of magnitude).

We employ the gravitational potential of \cite{PW80} to mimic the
effect of strong gravity near the NS surface.  In this potential the
Keplerian angular velocity takes the form 
\beq
\Omega_K^2=\frac{GM_*}{(R-R_g)^2R}.  
\eeq 
Note that the analytical work presented in the following sections 
is based on a Newtonian potential.

We specify the boundary conditions as follows.  We take the outer
boundary of the flow to be at $R_{\rm out}=10^6R_g$.  At this radius we
specify that the angular velocity is equal to its value in the
self-similar ADAF solution of \citet{NY94}, and that the proton and
electron temperatures are both equal to the self-similar ADAF
temperature.  We assume that the accreting star is a $1M_{\rm Sun}$
neutron star with a radius $R_*=3R_g=8.85$ km, unless stated otherwise.  
At $R=R_*$, we specify the value of the NS spin parameter,
$s=\Omega_*/\Omega_K(R_*)$, and we require the proton
temperature of the flow to be $T=\mathrm{few}\times10^7\textrm{ K} \ll
T_{virial}$.  (We do not assume that the electron and proton
temperatures are equal, but in fact they are equal.)  We do not
constrain the density of the gas in any way at either boundary.

The numerical problem as posed here has a family of solutions
characterized by three dimensionless parameters: the mass accretion
rate $\mdot$ (in Eddington units), the NS spin $s$ (in units of the
Keplerian angular velocity at the NS surface), and the viscosity
parameter $\alpha$.  The angular momentum flux $\dot J$, or
equivalently the dimensionless flux $j=\dot J/\dot
M\Omega_K(R_*)R_*^2$, is an eigenvalue of the problem.

\begin{figure}
\psfig{file=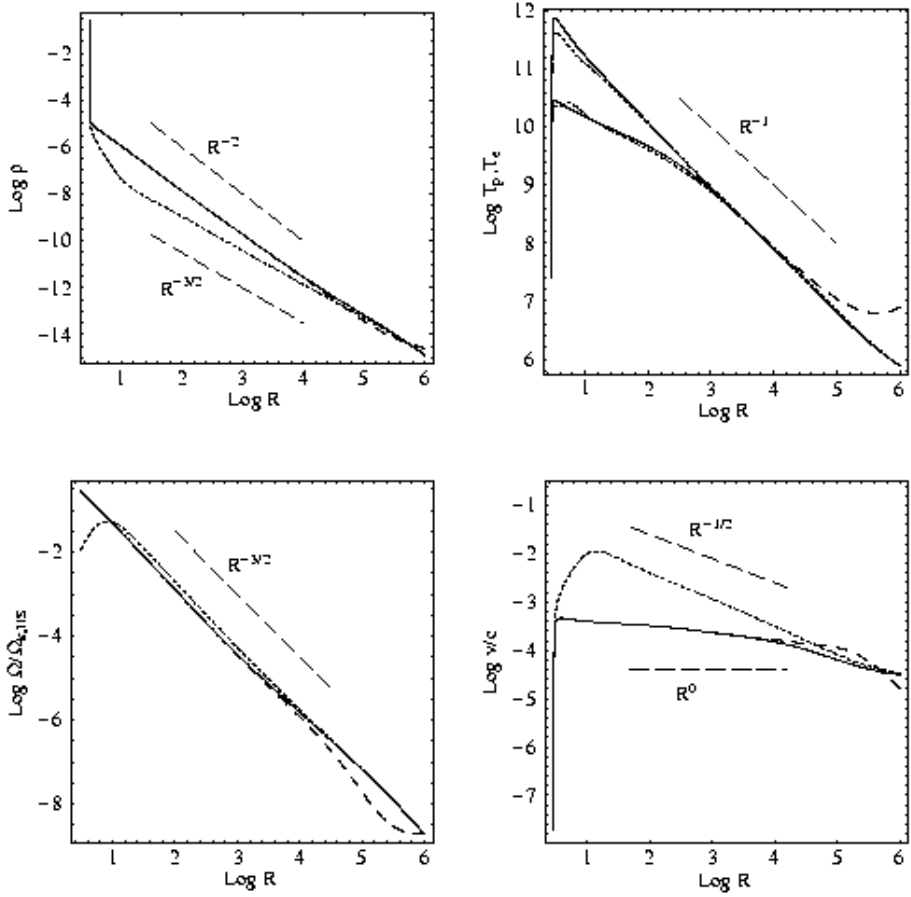,width=5in}
\caption{Profiles of density $\rho$ (g~cm$^{-3}$),
proton temperature $T_p$ ($^\circ$K), electron temperature $T_e$
($^\circ$K), angular velocity $\Omega$ (in units of the Keplerian
angular velocity at the NS radius $R_*$), and radial velocity $v$ (in
units of $c$) for accretion flows with $\alpha=0.1$, $\mdot=0.01$,
$\gamma=4/3$ and $s=0.3$ (solid curve) and $s=0.01$ (dotted curve).
The self-similar slopes for an ADAF flow and a settling flow
are shown for comparison.  The long-dashed curves represent the 
same solution as the solid curve, but with ten times higher 
temperature at $R_{out}$.
\label{f:ADAF} }
\end{figure}

It is  known that a hot flow with low $\dot m$ around a black hole is an 
``advection-dominated'' accretion flow (ADAF).  It has a two-temperature
structure for $R\la300R_g$ and is very hot (nearly virial) for
all $R$. This solution has indeed been obtained numerically
for $\dot m\la 0.01$ and $s\la 0.01$.
An interesting feature of the $s=0.01$ ADAF-type solution is that it
consists of two distinct segments.  For large radii 
(beyond $R\sim20R_g$ in Fig. \ref{f:ADAF}), the flow corresponds 
to the standard ADAF discussed in the literature, with the scalings 
\beq 
\rho\propto r^{-3/2}, \qquad c_s^2\propto r^{-1}, 
\qquad \Omega\propto r^{-3/2}, \qquad v\propto r^{-1/2}.  
\eeq 
However, at smaller radii, the
numerical solution indicates the presence of a second
advection-dominated zone, a ``settling ADAF,'' which was first seen in
numerical calculations described in \citet{NY94}.  This settling ADAF
is seen in Fig.\ \ref{f:ADAF} as a zone that lies between the
boundary layer region and the outer standard ADAF, with different
slopes for $\rho$ and $v$.  The radial extent of the settling ADAF
zone may be quite large and, in general, depends on the flow
parameters and boundary conditions.

A self-similar model of the settling ADAF may be readily obtained as
follows. In an ADAF, energy is not radiated, therefore $q^-=0$. Close
to the star $\Omega\simeq\textrm{ constant}$, therefore $q^+=0$.
Equations (\ref{energy-p}), (\ref{energy-e}) then simplify to the
condition of entropy conservation, $ds/dR=0$, which yields
$c_s^2\propto\rho^{\gamma-1}$.  As the material settles on the star,
its radial velocity decreases and we have $v\ll v_{ff},\
\Omega\ll\Omega_K$.  Then, from equation (\ref{mom}), it follows that
the temperature of the gas is nearly virial. Other quantities are
determined straightforwardly, so that we have 
\beq 
\rho\propto r^{\frac{1}{\gamma-1}}, \qquad c_s^2\propto r^{-1}, \qquad
\Omega\sim\textrm{const.}, \qquad v\propto r^{-\frac{2\gamma-3}{\gamma-1}}.  
\eeq 
The infall velocity decreases
with radius if $\gamma<1.5$, and increases if $\gamma>1.5$.  To
highlight the difference between the standard ADAF and the settling
ADAF, we have chosen $\gamma=4/3$ in the solutions shown in Fig.\
\ref{f:ADAF}.

\section{Hot brake flow}
\subsection{Numerical discovery}

By solving the equations numerically for different values of $s$, 
we have found that the with increasing $s$ the ADAF solution
continuously transforms into a solution of another type
 For relatively rapidly rotating
stars, with $s\ga0.1$, the new {\it hot break} solution (which was 
referred in the previous works as the {\it settling solution}) is 
already well-established. The transition is not sharp, so
it is difficult to identify a specific transition point $s=s_t$ at
which the transformation occurs.  Numerical experiments indicate that
the value of $s_t$ (however it is defined) is not very sensitive to
$R_{out}, \gamma$, and $\mdot$ and is, roughly, $s_t\sim0.04-0.06$.

The change of the nature of the flow as $s$ is varied is illustrated
in Fig.\ \ref{f:ADAF}.  The solid and dotted curves correspond to
two solutions with $s=0.3$ and $s=0.01$, respectively, with all other
boundary conditions being the same.  We see that the solutions are
markedly different from each other.  This is most clearly seen in the
profiles of density, where the $s=0.3$ model has a logarithmic slope
of $-2$, as appropriate for the cooling-dominated settling solution
described in this paper, and the $s=0.01$ model has a slope of $-3/2$,
as expected for a standard self-similar ADAF \citep{NY94,NY95a}.
There is a similar difference also in the profiles of the radial
velocity, where the two solutions have logarithmic slopes of $-1/2$ and
$0$, respectively.

\begin{figure}
\psfig{file=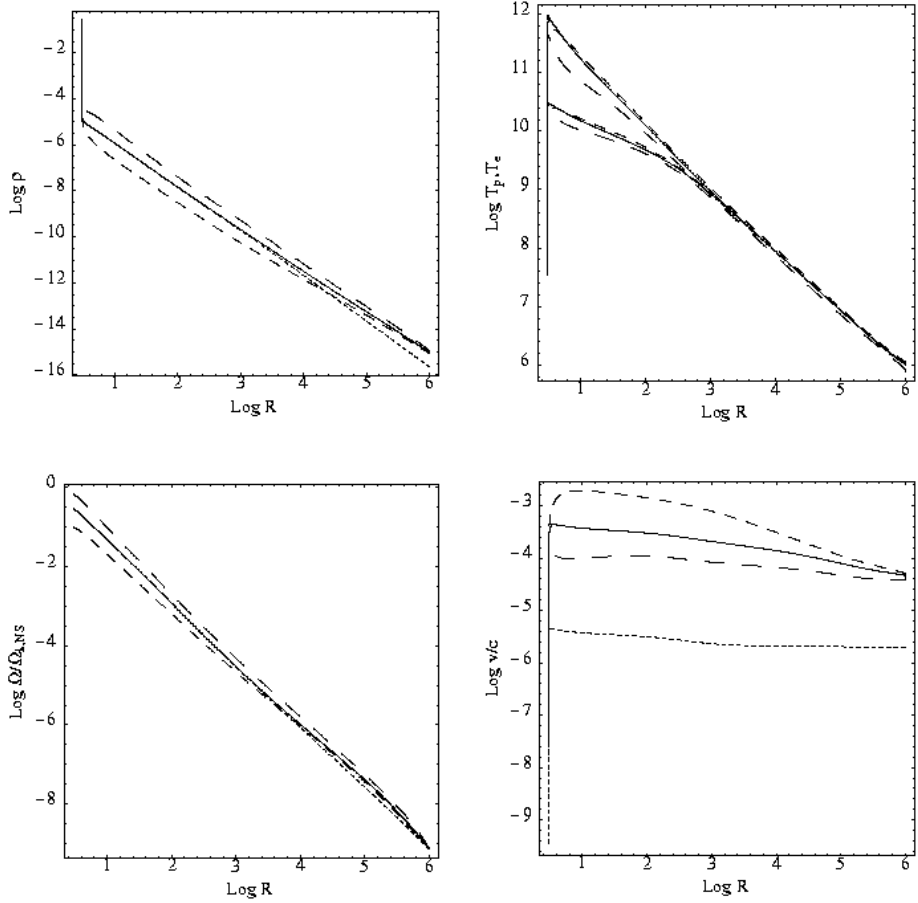,width=5in}
\caption{Profiles of density $\rho$ (g~cm$^{-3}$),
proton temperature $T_p$ ($^\circ$K), electron temperature $T_e$
($^\circ$K), angular velocity $\Omega$ (in units of the Keplerian
angular velocity at the NS radius $R_{NS}$), and radial velocity $v$ (in
units of $c$) for accretion flows with $\alpha=0.1$ and $(\mdot,s$) =
(0.01,0.3) -- solid line, (0.0001, 0.3) -- short-dashed line,
(0.01,0.1) -- medium-dashed line, (0.01,0.7) -- long-dashed line.
\label{f:BRAKE} }
\end{figure}

Figure \ref{f:BRAKE} shows representative solutions for $\alpha=0.1$
and a range of values of $\dot m$ and $s$.  The solutions clearly have
three radial zones.  For $R>300R_g$, there is a one-temperature zone
in which the gas properties vary roughly as power-laws of the radius.  For
$R<300R_g$, there is a second power-law zone with a two-temperature
structure.  Finally, close to the NS, the flow has a boundary layer
region.  In this final region, the gas experiences run-away cooling,
the velocity falls precipitously, and the density increases very
rapidly. 

The solutions are suspect also in the inner region of the two-temperature 
power-law zone and in the boundary layer, where Comptonization is likely 
to be important.  Outside these regions, however, the numerical solution
 is expected to be accurate.

We note here some qualitative results: (i) the boundary layer always
forms near the star surface, (ii) the transition radius where the
boundary layer meets the outer settling flow is usually at 
$R_{tr}\sim R_*$, (iii) the value of $R_{tr}$ depends on the 
NS spin. These results are in agreement with the
studies by \citet{TLM98} and \citet{TO99}, and are important for
the interpretation of kHz quasi-periodic oscillations.

An interesting property of the new solution is its insensitivity
to the value of accretion rate (provided it is low enough).
In our numerical solutions, curves corresponding to a given value
of $s$ and different values of $\mdot$ coincide with one other to very
good accuracy.  This is best
seen in the profiles of $\rho$ and $\Omega$.  Changing $s$ causes an
up/down shift of the curves but does not affect the slopes of the
curves. The temperature profiles are sensitive to the spin $s$,
especially for large values of $s$.  The radial velocity varies
approximately as $v\propto\mdot$ and is roughly consistent with
$v\propto r^0$ for $s>0.1$.

\subsection{Self-similar solution}
\label{1st}

We now find the hot break solution analytically. The hot brake flow
forms at low mass accretion rates. Therefore, for simplicity, 
we set $\dot m=0$ and omit the continuity equation.
Thus, the gas configuration corresponds to a radially static
``atmosphere.''  The motivation for this approximation follows from
the observation that the density $\rho$, temperature $T$ and the
angular velocity $\Omega$ of the gas in the \citet{MN01} self-similar solution
are completely independent of $\dot m$.  Only the radial velocity $v$
depends on $\dot m$, and it is given trivially by the spherical
continuity equation 
\beq 
v={\dot M\over 4\pi R^2\rho}.
\label{v}
\eeq 
Because of this, we do not lose any generality by setting $\dot
m=v=0$ in the analysis; we may always introduce a finite $\dot m$ and
finite $v$ after the fact. When $v \ne 0$, the energy
equation has an extra term corresponding to the advection of energy.
In advection-dominated accretion flows, for instance, this term
dominates over the cooling term $q^-$ \citep{NMQ97}.  In the present
case, however, we consider a situation in which the advection term is
negligible (which corresponds to low $\dot m$).

We assume that the flow is highly sub-Keplerian, $\Omega(R)
\ll\Omega_K(R)$, so that the centrifugal support is negligible
compared to the pressure support.  The radial momentum equation then
takes the following simple form, 
\beq 
{GM_*\over R^2}=-{1\over\rho}{d(\rho c_s^2)\over dR},
\label{mmtm}
\eeq 
where we have used the fact that $v\sim0$ and written the
pressure as $p=\rho c_s^2$ where $c_s$ is the isothermal sound speed.
\citet{MN01} present a more complete analysis in which they do not assume that
the rotation is slow.  They then obtain an extra factor of $(1-s^2)$
in their equation, which propagates through to all the results.  Since
we ignore the factor, our analysis corresponds to the case of a
slowly-spinning star: $s^2\ll1$.  This approximation is made only to
simplify the analysis, and all the results may be generalized for
arbitrary $s$.

From the analysis by \citet{MN01}, we know that the accreting gas in our
problem acts as a brake on the central spinning star and transports
angular momentum outward through the action of viscosity. We therefore
write the angular momentum conservation equation for the gas as follows, 
\beq 
\dot J=4\pi\nu\rho R^4{d\Omega\over dR}={\rm constant},
\label{jdot}
\eeq 
where $\dot J$ is the outward angular momentum flux, and $\nu$ is
the kinematic coefficient of viscosity. This equation is exactly valid
in steady state if $\dot m=0$. When $\dot m$ is non-zero, there is an
additional term, $\dot M\Omega R^2$, due to the flux of angular
momentum carried in by the accreting gas.  The key feature of the
hot brake solution is that the latter flux is negligible compared to
the outward flux from the star.  Equation (\ref{jdot}) is, therefore,
valid even when $\dot m\ne0$, so long as $\dot m$ is small enough for
the term $\dot M \Omega R^2$ to be negligible.

We first consider the {\em one-temperature} regime. The energy equations
for the electrons and protons reduce in this case to
\beq 
q^+\simeq q^- .
\label{energy}
\eeq 
The self-similar solution for the hot brake flow is then straightforwardly
obtained. The gas parameters have the following radial dependences: 
\beq
\rho=\rho_1r^{-2}, \quad T=T_1r^{-1}, \quad \Omega=\Omega_1r^{-3/2}.
\label{1st-sol}
\label{sss}
\eeq 
The subscript ``1'' in the coefficients is to indicate that this
is the first (hot brake) solution, to distinguish it from other
solutions described later.  By substituting the above solution in
equations (\ref{mmtm}), (\ref{jdot}) and (\ref{energy}), we see that
it satisfies the basic conservation laws.  We may also solve for the
numerical constants:
\begin{eqnarray}
\rho_1&=&\frac{\alpha s^2}{R_g}\,\frac{9}{2^{5/2}}
\left({\frac{m_e}{m_p}}\right)^{1/2}c^3Q_{\rm ff,NR},\\ kT_1&=&{m_p
c^2 \over 12},\\ \Omega_1&=&{s\,c \over \sqrt{2}R_g}=s\,\Omega_K(R_g).
\label{norm}
\end{eqnarray}
We note that if $\dot m \ne 0$ then the flow has a small
constant radial velocity: \beq v\propto r^0, \eeq as follows from
equation (\ref{v}).

The angular momentum flux in the solution is given by 
\beq 
\dot J=-\alpha^2 s^3 R_g^2\,\frac{3^{5/2}}{2^{5/2}}
\left({\frac{m_e}{m_p}}\right)^{1/2}\frac{c^5}{Q_{\rm ff,NR}}.
\label{jdot1}
\eeq 
By assumption, this flux is much greater than the angular
momentum flux due to accretion, which sets an upper limit on the mass
accretion rate for the solution to be valid \citep{MN01}.  

The pressure in the flow is given by 
\beq 
p =\rho c_s^2=\rho_1 c_{s1}^2\,r^{-3} \equiv p_1\,r^{-3},
\label{1st-true}
\eeq 
where $c_{s1}^2=2kT_1/m_p$. If the accretion flow is immersed into
an interstellar medium with some external gas pressure $p_{\rm ext}$,
then the above self-similar solution describes the flow at radii $r\ll
(p_1/p_{\rm ext})^{1/3}$, where the pressure $p \gg p_{\rm ext}$.  

This solution can be generalized to the {\em two-temperature} regime.
In this case the electrons and proton energies are governed by
two separate equations, which under our assumptions reduce to
\beq
q^+=q_{\rm Coul} ~\textrm{  and   }~ q_{\rm Coul}=q^-,
\eeq
respectively. The solution given by equations (\ref{1st-sol})
remains unchanged, with $T$ being the temperature 
of the proton component,  
\begin{eqnarray}
T_p&=&T_{p1}\,r^{-1},\\
T_{p1}&=&2T_1,
\end{eqnarray}
and the temperature of the electrons is
\begin{eqnarray}
T_e&=&T_{e1}\, r^{-1/2},\\
kT_{e1}&=&\left(\frac{Q_{\rm Coul}}{Q_{\rm ff,R}}\,
\frac{m_e^2 c^2}{m_p}\,kT_1\right)^{1/2}.
\end{eqnarray}

The obtained hot brake solution has the remarkable property that all
the quantities are uniquely determined by a single parameter $s$ ---
the dimensionless spin of the central object --- specified on the
inner boundary.  The fact that the solution does not depend on the
outer boundary condition in any way means that there is no simple way
to match it to the external medium.  Clearly, there has to be a second
solution to bridge the gap between this solution and the external
medium.  We derive the bridging solution now.

\section{Matching the hot brake flow with external medium}
\label{2nd}

We assume that the spinning star is immersed in a uniform
external medium with a density $\rho_{\rm ext}$, temperature $T_{\rm
ext}$ and pressure $p_{\rm ext}$.  We seek an accretion flow solution
that extends from the spinning star on the inside to the external
medium on the outside.  As we have seen above, there is the self-similar
hot brake solution extending from the boundary layer $R\sim R_*$ 
outward through a large distance, at least a few hundred $R_*$ or more. 
However, this self-similar solution has the surprising property
that the density, temperature and angular velocity of the gas at any
radius are completely independent of the outer boundary conditions.
Hence, the solution cannot be matched to a general external medium.
We resolve this paradoxical situation by showing that there is a
second self-similar solution which bridges the gap between the
original solution and the external medium (see \citealp{NM03}). 
This new solution has an
extra degree of freedom which permits it to match general outer
boundary conditions. 

We consider next the gas that lies just outside the region of validity
of the first self-similar solution described above.  In this zone, the
pressure is expected to be approximately equal to the external
pressure $p_{\rm ext}$: 
\beq 
\rho c_s^2 = p_{\rm ext} = {\rm constant}.
\label{mmtm2}
\eeq 
This condition replaces the hydrostatic equilibrium equation
(\ref{mmtm}), while equations (\ref{jdot}) and (\ref{energy}) continue
to be valid.  In this region, we find that there is a second
self-similar solution of the form \beq \rho=\rho_2r^{-7/2}, \quad
T=T_2r^{7/2}, \quad \Omega=\Omega_2r^{-9/4},
\label{2nd-sol}
\eeq where the label ``2'' refers to the fact that this is our second
solution.

To match the second and first solutions, we require that the fluxes of
angular momentum in the two solutions must be equal; this yields the
constraint
$(3/2)\rho_1\Omega_1T_1^{1/2}=(9/4)\rho_2\Omega_2T_2^{1/2}$.  
Making use of this and the other equations, we solve for the numerical
coefficients in equation (\ref{2nd-sol}):
\begin{eqnarray}
\rho_2&=&\frac{\alpha^{3/2} s^3}{p_{\rm ext}^{1/2}R_g^{3/2}}\,
\frac{3^{5/2}}{2^{9/4}}\left(\frac{m_e}{m_p}\right)^{3/4}
\frac{c^4}{Q_{\rm ff,NR}^{3/2}},\\
kT_2&=&\frac{p_{\rm ext}^{3/2}R_g^{3/2}}{\alpha^{3/2} s^3}\,
\frac{2^{5/4}}{3^{5/2}}\left(\frac{m_p}{m_e}\right)^{3/4}
\frac{m_p Q_{\rm ff,NR}^{3/2}}{c^4},\\
\Omega_2&=&\frac{\alpha^{1/4} s^{3/2}}{p_{\rm ext}^{1/4} R_g^{5/4}}\,
2^{-3/8}3^{-3/4}\left(\frac{m_e}{m_p}\right)^{1/8}
\frac{c^2}{Q_{\rm ff,NR}^{1/4}}.
\end{eqnarray}
The pressure in this solution is constant and equal to the external
pressure, $p_{\rm ext}$, and the angular momentum flux is also
constant and is equal to $\dot J$ in equation (\ref{jdot1}).  If the
flow has a small but nonzero accretion rate, $\dot m\not=0$, then its
radial velocity varies as [see eq. (\ref{v})] 
\beq 
v\propto r^{3/2}.
\eeq

Whereas the original hot brake self-similar solution has a unique profile
for a given choice of $s$, we see that the second solution derived
here has an extra degree of freedom, namely the external pressure
$p_{\rm ext}$. This extra degree of freedom solves the problem
discussed above.  Thus, the full solution consists of two zones: an
inner zone described by the first \citep{MN01} solution (\ref{1st-sol}) and
an outer zone described by the second solution (\ref{2nd-sol}).  The
radius $r_{\rm match}$ at which the two solutions match is obtained by
equating the pressures: \beq r_{\rm match}=\frac{\alpha^{1/3}
s^{2/3}}{p_{\rm ext}^{1/3} R_g^{1/3}}\, \frac{3^{1/3}}{2^{7/6}}
\left({\frac{m_e}{m_p}}\right)^{1/6}c^{5/3}Q_{\rm ff,NR}^{1/3}.
\label{R12}
\eeq

The second solution matches the external medium at the radius $r_{\rm
ext}$ at which its temperature matches that of the medium.  This gives
\beq r_{\rm ext}= \frac{\alpha^{3/7} s^{6/7}}{p_{\rm
ext}^{3/7}(kT_{\rm ext})^{2/7}R_g^{3/7}}\,
\frac{3^{5/7}}{2^{5/14}}\left(\frac{m_e}{m_p}\right)^{3/14}
\frac{c^{8/7}}{m_p Q_{\rm ff,NR}^{3/7}}.
\label{R23}
\eeq If we wish we could also write this in terms of the external
density by making the substitution 
$kT_{\rm ext} = m_p p_{\rm ext} /2\rho_{\rm ext}$.

\section{External medium solution}

For completeness, we present here the solution inside the external medium 
\citep{NM03}.  By assumption, the external medium has a uniform temperature
and density, and a uniform rate of cooling.  To maintain equilibrium,
there has to be some constant source of heat that exactly compensates
for the cooling.  We assume that such a source of heat exists (e.g.,
cosmic rays).  The rotation $\Omega$ is non-zero, but it decays
rapidly outward.  The small amount of rotation helps to transport the
angular momentum flux from the star out into the external medium.
Solving the angular momentum conservation law (3), we obtain the
following solution \beq \rho=\rho_{\rm ext}, \quad T=T_{\rm ext},
\quad \Omega=\Omega_3 r^{-4},
\label{3rd-sol}
\eeq where \beq \Omega_3 = \alpha\, s^3\,\frac{3^{5/2}}{2^7\pi}
\frac{m_e^{1/2} c^5}{R_g^2 Q_{\rm ff,NR}}\, \rho_{\rm ext}^{-1}\left(k
T_{\rm ext}\right)^{-1/2} , \eeq and $p_{\rm ext}=2kT_{\rm
ext}\rho_{\rm ext}/m_p$.  For $\dot m\not=0$, the velocity scales as
$r^{-2}$.

\begin{figure}
\psfig{file=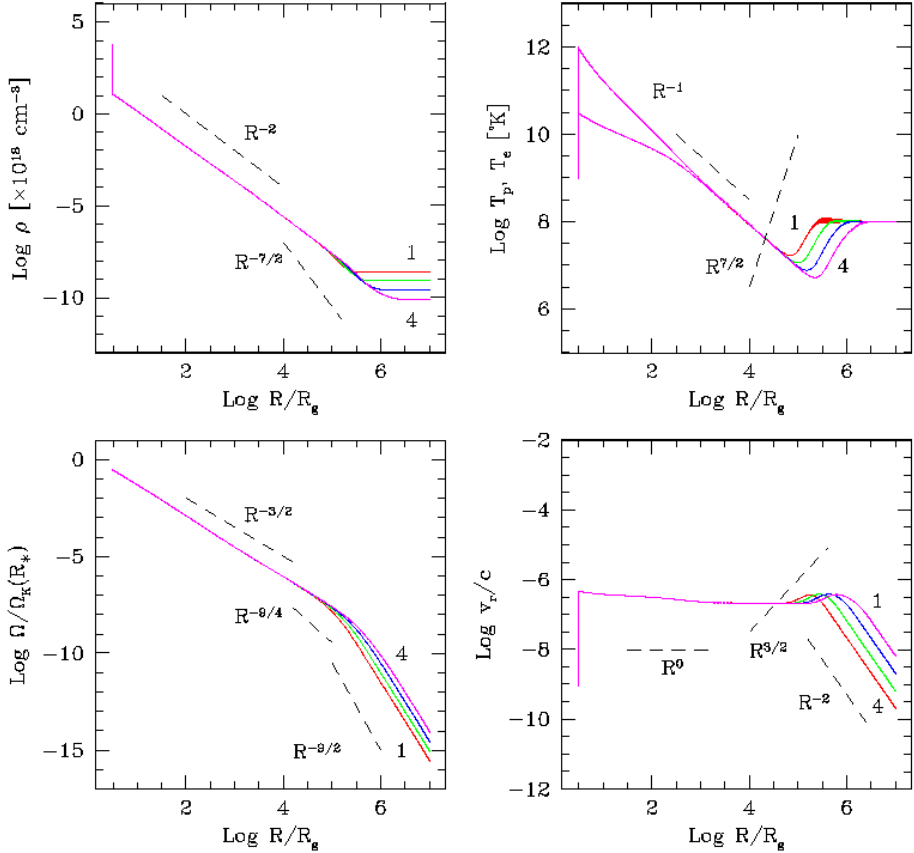,width=5in}
\caption{Profiles of density (top left panel), temperature (top right
panel, the electron temperature is the lower curve on the left and the
proton temperature is the higher curve), angular velocity (bottom left
panel), and radial velocity (bottom right panel), for four numerical
solutions of the full height-integrated differential equations.  The
four solutions correspond to different values of the density of the
external medium: $\rho_{\rm ext}= 2.5\times10^9,\ 8.1\times10^8,\
2.5\times10^8,\ 8.1\times10^7$~cm$^{-3}$.  The first and fourth
solutions are labeled 1 and 4, respectively.  The temperature of the
external medium and the accretion rate are kept fixed in all the
solutions: $T_{p,{\rm ext}}=T_{e,{\rm ext}}=10^8$K, $\dot m=2\times
10^{-5}$.  The analytical slopes of the three self-similar solutions
are shown for comparison
\label{f:2nd}}
\end{figure}

We confirm the existence and the structure of our matching solution 
(and the external medium) using the same numerical model \citep{NM03}. 
In the calculations, the flow was taken to extend from an inner radius
$R_{\rm in}=3~R_g$ to $R_{\rm out}=10^7~R_g$.  The mass accretion rate
was taken to be low, $\dot m=2\times 10^{-5}$, in order that the flow
should correspond to the regime of the hot settling flow solution.  We
took the viscosity parameter to be $\alpha=0.1$ and set the spin of
the star to be $s=0.3$ (i.e., 30\% of the Keplerian rotation at the
stellar surface).  We took the other inner boundary conditions to be
the same as in MN01.  At the outer boundary, we specified the
temperature and density of the external medium.  Figure \ref{f:2nd}
shows four solutions.  The external temperature is kept fixed at
$T(R_{\rm ext})=10^8$~K in all the solutions, but the external density
varies by a decade and a half: $\rho(R_{\rm ext})= 2.5\times10^9,\
8.1\times10^8,\ 2.5\times10^8,\ 8.1\times10^7$~cm$^{-3}$.  We have
also done other calculations in which we kept $\rho_{\rm ext}$ fixed
and varied $T_{\rm ext}$.  These give very similar results.

Fig. \ref{f:2nd} shows that, right next to the star, there is a
boundary layer, where the density rises sharply as one goes into the
star and the temperature drops suddenly.  We do not analyze this
region.  Once we are outside the boundary layer, the gas behaves very
much according to the analytical solutions discussed in Section \ref{1st}.  
Starting just outside the boundary layer and extending over a wide range of
radius, the numerical solution exhibits a self-similar behavior with
power-law dependences of the density, temperature and angular
velocity.  This region corresponds to the self-similar solution of
MN01.  There are, in fact, two zones, an inner two-temperature zone,
and an outer one-temperature zone \citep{MN01} [see also Eq. (\ref{1st-sol})].
The most notable feature of this region is that the density, temperature 
and angular velocity of the numerical solutions are completely independent 
of the outer temperature and density, as predicted by the analytical
solution.  The slopes of the numerical curves also agree well with the
analytical scalings.

At a radius $R_{\rm match}\sim5\times10^4 ... 2\times10^5~R_g$
[depending on the outer pressure, see Eq. (\ref{R12})], solution 1
merges with solution 2 [Eq. (\ref{2nd-sol})] described in the previous section.
Here, the solution does depend on the outer boundary conditions, and it
scales roughly according to the slopes derived analytically.  At even
larger radii $R > R_{\rm ext}\sim3\times10^5 ... 2\times10^6~R_g$ [see
Eq. (\ref{R23})], the flow matches onto the ambient external medium.  In
this region we have solution 3 [Eq. (\ref{3rd-sol})] described in the present 
section. As expected, out here only the angular velocity and the radial
velocity vary with radius.  Both have the scalings predicted for
solution 3.

\section{Boundary layer solution}  

Finally we discuss the last piece: the structure of the boundary layer.
Unlike the previous cases, the solution to the boundary layer (BL) cannot
be obtained in a self-similar form in terms of the radial coordinate $R$.
The structure of the BL is intrinsically non-self-similar in $R$ since
all the gas parameters (e.g., the temperature, gas density, etc.) change
very dramatically over a relatively short radial region: 
$R_*\le R \la 2 R_*$, as can be seen from the numerical solutions 
discussed in the previous sections. For instance, the density nearly
diverges as one gets close to the star surface whereas the temperature
decreases to the values well below the virial temperature.  
Such a behavior, however, suggests to look for a self-similar solution
in terms of the distance from the stellar surface, i.e., in terms
of $D=R-R_*$. In this calculation we neglect the effects of radiation
transfer and, especially, the Comptonization.
They may be important in hot regions, but will unlikely strongly
affect the flow closer to the star, where the temperature of the
gas falls below few$\times10^{9}$~K or so (see discussion in 
Section \ref{comptoniz}). 

Unlike all previous cases, here we cannot neglect the radial (infall)
velocity. For simplicity, we consider the one-temperature 
case. The generalization to the two-temperature case is straightforward:
it follows from the equality of the Coulomb energy transfer rate and
the heating/cooling rates (see Section \ref{1st}) and will be discussed
in a separate paper.  
We again use the height-integrated hydrodynamic equations, but now 
written in the approximation that $R=R_*+D$ with $D\ll R_*$~:
\bea
& \displaystyle -\dot M=4\pi R_* \rho v, & \label{1}\\
& \displaystyle v\frac{dv}{dD}=\left(\Omega^2-\Omega_{K*}^2\right)R_* -
\frac{1}{\rho}\,\frac{d}{dD}\left(\rho c_s^2\right), & \label{2}\\
& \displaystyle 4\pi\alpha\left(\rho c_s^2\right)\frac{R_*^4}{\Omega_{K*}}\,
\frac{d\Omega}{dD}=\dot J-\dot M\Omega R_*^2, & \label{3}\\
& \displaystyle \frac{\rho v}{\gamma-1}\,\frac{d c_s^2}{dD}
-c_s^2 v\frac{d\rho}{dD}=q^+-q^-, & \label{4}
\eea
where
\bea
q^+&=&\alpha\left(\rho c_s^2\right)\,\frac{R_*^2}{\Omega_{K*}}
\left(\frac{d\Omega}{dD}\right)^2,\\
q^-&=&Q_{\rm ff,NR}\rho^2\sqrt{c_s^2},
\eea
and in equation (\ref{1}) we took into account that the radial velocity
is negative (inward).

As we mentioned above, we are looking in the solution which is self-similar 
in $D$, that is the temperature, density, angular and radial velocities
are expressed as power-laws. In addition, we must satisfy the boundary 
condition at the star surface: $\Omega=\Omega_*$. Thus, we readily conclude 
that $\Omega\propto D^0$ (otherwise it is either zero or diverges 
at $D=0$, i.e., $R=R_*$).

Let us now consider equation (\ref{2}). First, we note that the rotation
is sub-Keplerian, $\Omega_*^2\ll \Omega_{K*}^2$ so that we neglect 
this term. Next, we cast it into the form:
\beq
\frac{d}{dD}\left(c_s^2+\frac{1}{2}v^2 \right)+
c_s^2\left(\frac{1}{\rho}\,\frac{d\rho}{dD}
+\frac{1}{2}\,\frac{v_{\rm ff,*}}{c_s^2} \right)=0,
\label{momnt}
\eeq
where $v_{\rm ff,*}=\sqrt{2}\Omega_{k*}R_*$ is the free-fall velocity 
that near the stellar surface. We now make the following assumptions,
which consistency with the obtained solution must be checked
{\it a posteriori}: (i) the flow is always subsonic, $v^2\ll c_s^2$, and
(ii) $c_s^2$ grows with $D$ slower than linearly 
(for $c_s^2\propto D^{2\beta}$ we should have $2\beta<1$), then
the second term in the second brackets is sub-dominant and may be 
neglected as well. With these assumptions, the equation simplifies to
\beq
\rho c_s^2 = p = \textrm{ const.},
\label{pres}
\eeq
that is, the pressure is constant throughout the boundary layer.

Considering equation (\ref{3}), we notice that since $\Omega=$const.,
its radial derivative vanishes, $d\Omega/dD=0$, and the equation simply
defines the angular momentum flux, $\dot J=\dot M\Omega_* R_*^2$.
By the same token, the heating rate in equation (\ref{4}) vanishes. 
Together with the continuity equation (\ref{1}), the energy equation reads,
\beq
\frac{\dot M}{4\pi R_*^2}\left(\frac{1}{\gamma-1}\,\frac{d\,c_s^2}{dD}
-\frac{c_s^2}{\rho}\,\frac{d\rho}{dD}\right)=Q_{\rm ff,NR}\rho^2 c_s
\label{ener}
\eeq

The system of equations (\ref{1}), (\ref{pres}), (\ref{ener}), together
with $\Omega=\Omega_*$ admits the following self-similar solution:
\beq
\rho=\rho_0\, d^{-2/5},\qquad T=T_0\, d^{2/5},
\qquad v=v_0\, d^{2/5},\qquad \Omega=\Omega_0\, d^0,
\label{BLsol}
\eeq
where we used the dimensionless distance $d=D/R_*$. The constant 
factors are (recall that $c_s^2=kT/m_p$):
\beq
\rho_0=\frac{p_{\rm out}}{B^2}, \qquad
kT_0=m_p\,B^2, \qquad
v_0=\frac{\dot M}{4\pi R_*^2\rho_0}, \qquad
\Omega_0=\Omega_*,
\eeq
where we denoted
\beq
B=10\pi\, Q_{\rm ff,NR}\,\left(\frac{\gamma}{\gamma-1}\right)\,
\frac{R_*^3}{\dot M}
\eeq
and $p_{\rm out}$ is the pressure on the outside of the boundary layer,
where it should match to the pressure on the inside of the hot brake flow.
Using equations of section \ref{1st}, we can calculate this pressure to be
equal to
\beq
p_{\rm out}=\frac{3}{16\sqrt{2}}\, (\alpha s^2)\,
\frac{Q_{\rm ff,NR} c^5}{R_g}\,\left(\frac{R_g}{R_*}\right)^3\,
\left(\frac{m_e}{m_p}\right)^{1/2}.
\eeq 
With the solution we check that the assumptions made in order to 
simplify equation (\ref{momnt}) are consistent: indeed 
$v^2/c_s^2\propto d^{2/5}\to 0$ as $d\to 0$ and $2\beta=2/5<1$.
There is a little subtlety with Eq. (\ref{momnt}), namely the 
condition $\rho c_s^2=$const. ensures that the two leading terms
cancel each other exactly and it looks like we need to keep higher order
terms, e.g., those that have been dropped out. Apparently, it is 
important to understand that the presented self-similar solution is
approximate: the flow is pressure-dominated but with 
$\rho c_s^2\approx$constant. A small deviation from the exact 
equality is necessary to compensate for the remaining next order terms.  

\begin{figure}
\psfig{file=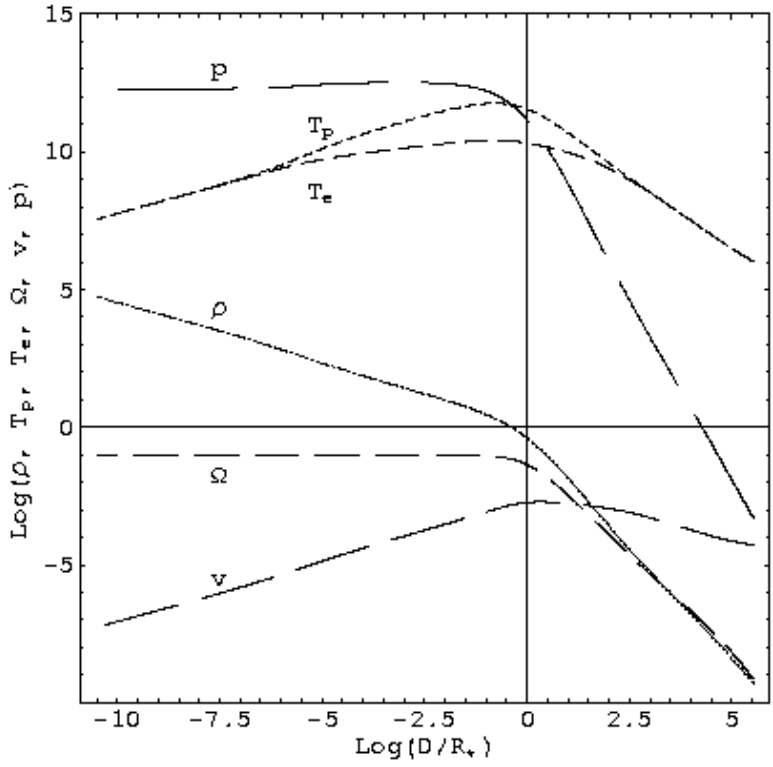,width=4in}
\caption{The structure of the boundary layer. $Log$ of $\rho,~ T_e,~ T_p,~
\Omega,~ v$~ and $p=\rho c_s^2$ as functions of $Log(D/R_*)$ are shown
for $\dot m=0.01,~ \alpha=0.1,~ s=0.1$.
\label{f:BL} }
\end{figure}

Using our computer code, we calculated the structure of the boundary 
layer numerically and presented it in Figure \ref{f:BL}.
Note the remarkable agreement of this numerical solution with the 
theoretical one: $\Omega=$const., $p\approx$const., and 
$\rho,~ T_p,~ v~$ follow the predicted scalings.

\section{Properties of the hot brake accretion flow}
\subsection{Stability of the flow}

The existence of a mathematical solution to hydrodynamic equations
does not imply that the corresponding accretion flow may realize in nature.
If the flow is unstable to a certain
type of instability, this instability may dramatically change the 
structure of the entire flow, or even prohibit it from being realized 
in nature. Here we perform the linear stability analysis of the 
hot brake flow.

\subsubsection{Stability to winds/outflows}

It is known that the Bernoulli parameter of the accreting gas in BH
ADAFs is positive for a wide range of $r$ \citep{NY94,NY95a,NKH97}, 
and it has been suggested that the positive Bernoulli parameter may trigger 
strong winds or jets in these systems \citetext{\citealp{NY94,NY95a,BB99},
but see \citealp{ALI00}}.
\citet{IA99,IA00} confirmed with numerical simulations that strong 
outflows are produced from BH ADAFs when $\alpha\sim1$.

Normalizing the Bernoulli parameter, $Be$, by $(\Omega_KR)^2$,
and using equations (\ref{sss}),(\ref{norm}),
we find that the self-similar hot brake flow has
\bea
b\equiv {Be\over\Omega_K^2R^2}
&=&\frac{1}{v_K^2}\left(\frac{1}{2}v^2+\frac{1}{2}\Omega^2R^2-\Omega_K^2R^2
+\frac{\gamma}{\gamma-1}c_s^2\right) 
\nonumber\\
&\simeq&-\frac{2\gamma-3}{3(\gamma-1)},
\label{Bern}
\eea
where $\gamma$ is the mean adiabatic index of the flow.

The right hand side of equation (\ref{Bern}) can be either
positive or negative, depending on the value of $\gamma$. Hence, 
we find that the gas is gravitationally bound and
unable to flow out in a wind (i.e. $b<0$) if the adiabatic index
satisfies 
\beq 
\gamma>\frac{3}{2};
\label{b<0}
\eeq 
that is, the accretion 
flow can produce a wind and/or a collimated outflow only if $\gamma$ 
is well below that of an ideal has ($\gamma=5/3$) 
and is stable to such outflows if $\gamma>1.5$.
Normally, we expect $\gamma$ to be close to 5/3 for the accreting gas.

\subsubsection{Stability to convection}

It is well known that if the entropy increases inward in a
gravitationally-bound non-rotating system, the gas is
convectively unstable; otherwise the flow is stable.  The specific
entropy profile in the settling accretion flow around a NS can be readily
calculated from equations (\ref{energy-p}),(\ref{energy-e}) using
(\ref{sss}),(\ref{norm}).  This gives 
\beq
\frac{ds}{dR}=\frac{k}{m_p}\frac{1}{\gamma-1}\frac{d}{dR}
\ln\left(\frac{c_s^2}{\rho^{\gamma-1}}\right)
=\frac{k}{m_p}\frac{2\gamma-3}{\gamma-1}\frac{1}{R}.  
\eeq 
We see that the entropy
increases outward for $\gamma>1.5$ and inward for $\gamma<1.5$.
Hence if $\gamma>1.5$ the flow is stable against convection, while if
$\gamma<1.5$ the flow is convectively unstable.

In the presence of rotation, the analysis is a little more complicated.
\citet{NIA99} and \citet{QG99} discuss the generalization of the 
Schwarzchild criterion for accretion flows with rotation. 
If the gas motions are restricted to the equatorial plane
of a height-integrated flow, convective 
stability requires the following effective frequency to be positive:
\beq
N^2_{\rm eff}=N^2+\kappa^2>0,
\eeq
where $N$ is the Brunt-V\"ais\"al\"a frequency and $\kappa$ is the epicyclic 
frequency, $\kappa=\Omega$ for $\Omega\propto R^{-3/2}$. For a power-law
flow with $\rho\propto R^{-a}$ and $\Omega(R)=s\Omega_K\propto R^{-3/2}$
with $s^2=1-(1+a)c_0^2$, this criterion may be written as follows
\citetext{see \citealp{NIA99} for more discussion}
\beq
N^2_{\rm eff}=\Omega^2_K\left(-[(\gamma+1)-a(\gamma-1)]
\frac{(1+a)c_0^2}{\gamma}+1\right)>0.
\label{Neff}
\eeq
Since for the self-similar settling solution $a=2$, the stability criterion 
(\ref{Neff}) becomes
\beq
N_{\rm eff}^2\approx\frac{\Omega_K^2}{\gamma}
\,(2\gamma-3)>0
\eeq
which yields that the flow is convectively stable if
\beq
\gamma>\frac{3}{2}.
\label{c>0}
\eeq 
This condition is different from the stability criterion against outflows,
given in equation (\ref{b<0}).

Following the techniques developed by \citet{QG99}, \citet{NIA99} have
also presented a more general analysis of convection in a self-similar
accretion flow.  This analysis, which does not restrict motions to lie
in the equatorial plane, assumes that $v_\phi$ and $c_s$ are
independent of the polar angle $\theta$ \citetext{as is valid for a marginally
convectively stable system, cf \citealp{QG99}}.  \citet{NIA99} find
that the most unstable region of the flow is near the rotation axis,
$\theta=0,\pi$.  They show that the marginal stability criterion for
this polar fluid coincides with the condition for the positivity of
the Bernoulli parameter.  That is, a flow which is convectively stable
at all $\theta$ has a negative Bernoulli parameter, while a flow which
is convectively unstable for at least some values of $\theta$ has a
positive Bernoulli parameter.  (The Bernoulli parameter itself is
independent of $\theta$.)

We have verified this result for the solutions presented in this
paper.  Specifically, when we apply to our solution the more general convective
stability criterion given by equation (A9) of \citet{NIA99}, 
we recover the condition (\ref{b<0}) above.

\subsubsection{Thermal stability}

Not all hot accretion flows are stable.  For instance, the cooling-dominated 
SLE solution has been shown to be thermally unstable
\citep{P78,WL91,NY95b} and, hence, unlikely to exist in nature.  More
generally, it has been shown that any accretion flow in which heating
balances cooling is thermally unstable if the cooling is due to
bremsstrahlung emission \citep{SS76,P78}. The ADAF solution, on the
other hand, is known to be thermally stable
\citep{NY95b,K+96,K+97}.  In this solution, cooling is weak (ideally
zero), and so the thermal energy of the flow is not radiated but is
advected with the gas (hence the name). The CDAF is also believed to
be stable, since in this flow again the thermal energy is advected by
convective eddies and is either carried into the black hole or is
radiated near the outer boundary of the flow \citep{BNQ01}.
In contrast to these radiatively inefficient flow, the hot brake flow
is cooling-dominated. Energetically, this flow is very similar to
the SLE solution, since the heat energy produced by viscous
dissipation is radiated locally via bremsstrahlung.  Therefore, in
analogy with the SLE solution, one might expect the flow to be
thermally unstable.  However, this is not necessarily the case, as we
show in this section (see also, \citealp{MN02}).

The physics of the thermal instability is simple \citep{F65}.  Suppose
a system is in thermal equilibrium, so that the rates of heating and
cooling per unit volume are equal: $Q^+=Q^-$.  For simplicity let us
take the heating and cooling rates to be functions of only the local
temperature: $Q^+\propto T^\alpha$, $Q^-\propto T^\beta$ ($\alpha,\
\beta>0$ for concreteness).

Suppose, with increasing temperature, the cooling rate rises faster
than the heating rate, i.e., $\beta>\alpha$. Then a local perturbation
which causes a small increase in the temperature will result in a net
cooling of the gas: $Q^->Q^+$. This will cause the temperature to
return to its equilibrium value, which means that the gas will be
thermally stable.  (It is easily seen that this is true also for a
small decrease in the temperature.)  On the other hand, if
$\alpha>\beta$, the gas is thermally unstable. For instance, if the
temperature decreases slightly, cooling becomes stronger than heating
and the system deviates from its equilibrium in a run-away manner.

When a gas is hot, one should always include the effect of thermal conduction,
--- being  usually a strong function of temperature, thermal conduction
may (and will) affect the accretion flow structure and, especially, its
stability. 

It is easy to see that thermal conduction will tend to reduce the
thermal instability.  An unstable thermal mode of wave-vector $k$
consists of a growing temperature perturbation of wave-length
$2\pi/k$.  Thermal conduction tends to smooth out this temperature
perturbation through heat diffusion.  If the rate at which the
temperature perturbation grows is smaller than the rate at which it is
smoothed out by conduction, then the instability will be suppressed
and the mode will be stable. Otherwise, the mode will continue to
grow, but at a somewhat reduced rate.  

The rate at which fluctuations are smoothed out by conduction depends
on the spatial scale of the perturbation. The smaller the scale
(i.e. the larger the value of $k$), the faster the conduction, and the
greater the stabilizing effect.  Thus, we expect conduction to
stabilize thermal modes with $k$ greater than some critical $k_{\rm
crit}$.  Our task in this section is to
estimate $k_{\rm crit}$ through a quantitative analysis.  If we find
that $k_{\rm crit}R\gg1$, then we conclude that the flow is thermally
unstable.  On the other hand, if we find that $k_{\rm crit}R \lesssim
1$, we may reasonably claim that the flow is thermally stable.
Technically, for $k\sim1/R$, we need to carry out a global analysis
rather than the local analysis presented in this paper, but this is
beyond the scope of the present section.

Let us write  the heat flux $q$ due to thermal conduction as
\beq 
q_{\rm cond}=-\kappa\nabla T,
\label{q-grad}
\eeq
where $\kappa$ is the thermal conductivity coefficient. Thermal 
conductivity in a dense, fully ionized gas is given by the
\citet{Spitzer62} formula,
\beq
\kappa_{\rm Sp}\approx1.3nk_Bv_T\lambda 
\simeq6.2\times10^{-7}T_e^{5/2}\textrm{ erg/(s K cm)}.
\label{kappa-Sp}
\eeq 
Here $v_T=(k_BT_e/m_e)^{1/2}$ is the electron thermal speed,
$T_e$ is the electron temperature ($T_e=T$ for a one-temperature
plasma), $k_B$ is the Boltzmann constant, and 
\beq 
\lambda\simeq10^4 T_e^2/n\textrm{ cm}
\label{mfp}
\eeq 
is the electron mean free path.  Note that $\lambda$ is
independent of the mass of the particle.

In the collisionless regime, i.e., when the mean free path of an
electron becomes comparable to or larger than the temperature gradient
scale $\lambda\ga T_e/|\nabla T_e|$, equation (\ref{q-grad}) for the
heat flux is no longer valid.  For an unmagnetized plasma, the heat
flux takes the following saturated form \citep{CMc77}, 
\beq 
q_{\rm sat}\simeq-C\rho c_s^3\textrm{ sgn}(\nabla T),
\label{q-sat}
\eeq 
where $C\sim5$ is a numerical constant whose exact value depends
on the particle distribution function.  This result is not relevant
for our problem since our plasma is magnetized.

For a collisionless magnetized plasma, thermal conduction is
anisotropic.  Electrons stream freely along the field lines, and the
parallel heat flux remains the same as for the unmagnetized case
described above.  However, the transverse heat flux is greatly reduced
because electrons are tied to the field lines on the scale of the
Larmor orbit.  In fact, if the field is uniform and homogeneous, the
perpendicular thermal flux is identically equal to zero since
electrons cannot move across the field lines.  In a tangled field,
however, electrons can jump from one field line to another and thus
conduct heat perpendicular to the field.  Since we are dealing with a
turbulent accretion flow with a tangled magnetic field, this is the
regime of interest to us.

The physics of this regime of conduction has been discussed by
\citet{RR78,CC98,MN02}, who identified two important effects,
which we discuss now. 
 
First, since particles can move freely only along field lines, the
characteristic effective mean free path is set by the correlation
scale of the magnetic field $l_B$.  In a hot accretion flow this scale
is not known in general. However, it is likely that turbulent motions
in the flow occur on a scale comparable to the local radius $R$, since
this is the only characteristic scale in the problem.  Very likely,
the turbulent magnetic field will also have the same scale $l_B\sim
R$.  We parameterize this scale as $l_B=\xi R$.  We expect $\xi\le1$
because turbulent fluctuations cannot have a scale larger than the
local radius of the flow. We assume $\xi\sim0.1$ throughout the paper.

Second, the magnetic field is inhomogeneous.  Therefore, only a
fraction $\vartheta<1$ of the particles will be able to pass though
the magnetic mirrors that will be present in the field, and it is only
these particles that transport energy beyond a distance $\sim l_B$.
For magnetic field strength fluctuations $\delta B\sim\langle
B\rangle$, the fraction of free streaming particles is estimated to be
$\vartheta\sim0.3$.

Typically, hot accretion flows are highly collisionless, i.e.,
$\lambda\gg R\ga l_B$.  Therefore, we can write the thermal conduction
coefficient as 
\beq 
\kappa_B\simeq nk_Bv_Tl_B\,\vartheta \simeq
10^{-2}nk_Bv_TR\xi_{-1}\vartheta_{-1},
\label{kappa_B-am}
\eeq 
where $\xi_{-1}=\xi/10^{-1}$ and
$\vartheta_{-1}=\vartheta/10^{-1}$.  Let us write the conductive heat
flux in a form similar to that used for the viscous stress, namely
\beq 
q_{\rm cond}=-\alpha_c\frac{c_s^2}{\Omega_K}\rho\frac{dc_s^2}{dx},
\label{q-cond}
\eeq 
where the dimensionless coefficient $\alpha_c$ is analogous to
the Shakura-Sunyaev viscosity parameter $\alpha$, and is given by 
\beq
\alpha_c \simeq\frac{R}{H}\,\xi\vartheta
\simeq10^{-2}\xi_{-1}\vartheta_{-1}.
\label{alpha_c}
\eeq 
Here we have used the fact that $v_T\simeq c_{se}$ and $H/R\sim
c_s/v_{\rm ff} \sim c_s/\Omega_KR$, where $H$ is the accretion disk
scale height (in hot flows, $H\sim R$) and $v_{\rm ff}$ is the
free-fall speed.  

To study the thermal stability of an accretion flow, we need to
include additional physics, namely the effects of shear and rotation.
We use the shearing sheet approximation \citep{GLb65,JT66,GT78}, which
is a convenient way of introducing the relevant physics without
unnecessary technical complications.  Furthermore, we carry out a
local WKB analysis under the assumption that the wavelength of the
perturbation is much smaller than the radius.

The shearing sheet model approximates the flow as locally flat,
neglecting the effects of the flow curvature.  Conventionally, the
shearing sheet coordinates are Cartesian with $x,\ y,\ z$
corresponding to the radial, azimuthal, and vertical directions,
respectively, centered on some point in the flow at a radius
$R$. These coordinates are appropriate for describing the motion of a
parcel of gas whose geometrical size is small compared to the local
radius, $R$, of the flow (i.e., $x,\ y,\ z\ll R$), so that the effects
of geometry and curvature are insignificant.  We neglect viscosity in
the azimuthal momentum equation; of course, we do include viscous
dissipation in the energy equation, where it plays an important role.

It is convenient to compare the wave-vector $k$ of a perturbation with
$1/R$ and the frequency of a mode with the local Keplerian frequency
$\Omega_K=\sqrt{GM_*/R^3}$, where $M$ is the mass of the central object.
The shearing sheet approximation is accurate for ``local'' small-scale
perturbations with $kR\gg1$.  Perturbations with $kR\sim1$ are global;
their properties may be understood only through a global stability
analysis, usually numerical, which we do not attempt here.

We consider a shearing gas flow with unperturbed velocity given by
\beq {\bf V}_0(x)=2A\,x\,\hat{y}, \eeq where $2A=d{\bf V}_0/dx$ is the
shear frequency and ``hat'' denotes a unit vector.  Note that we have
neglected the radial velocity in the equilibrium flow since this
component of the velocity is significantly smaller than the azimuthal
velocity.  To
include the effect of rotation we assume that there is a Coriolis
acceleration, described by an angular rotation frequency 
${\bf \Omega}=\Omega\,\hat{z}$. The vorticity and epicyclic 
frequency are then given by 
\beq 
2B=2A+2\Omega,\qquad \kappa_{\rm epi}=2(\Omega B)^{1/2}.  
\eeq 
The hot brake flow
satisfies the Keplerian scaling, $\Omega\propto R^{-3/2}$.  Therefore,
we have $2A=-(3/2)\Omega$, $2B=\Omega/2$ and
$\kappa_{\rm epi}=\Omega$.

We assume that perturbations in the flow have structure only in the
$x$ direction, and we ignore motions in the $z$ direction.  We write
the perturbations (represented by primes) in the velocity, density and
sound speed as 
\bea 
{\bf V}'(x,t)&=&u(x,t)\,\hat{x}+v(x,t)\,\hat{y},\\
\rho'(x,t)&=&\rho_0\sigma(x,t), \\ 
{c_s^2}'(x,t)&=&a^2(x,t), 
\eea
where $\rho_0$ and $c_s^2$ are the equilibrium values of the density
and the square of the sound speed.  Note that we define $c_s$ to be
the isothermal sound speed, so that the pressure is written as 
$p=\rho c_s^2$.  By considering perturbations of the basic hydrodynamic
equations, namely the continuity, radial momentum, azimuthal momentum
and entropy equations, we obtain the following four linearized equations
(note that in the presence of conduction the energy equation has 
an additional contribution from the divergence of $q_{\rm cond}$, 
and in equilibrium, the total heating is equal to the total cooling),
\bea 
& &\dt{\sigma}+\dx{u}=0,\label{eqs}\\ 
& &\dt{u}-2\Omega v+c_s^2\dx{\sigma}+\dx{a^2}=0,\\ & &\dt{v}+2Bu=0,\\ 
& &\frac{\rho_0}{\gamma-1}\dt{a^2}-\rho_0 c_s^2\dt{\sigma}
=\left(Q^++Q^-\right)'+\alpha_c\frac{\rho c_s^2}{\Omega_K}
\frac{\partial^2 a^2}{\partial x^2},
\eea
where we have used $d/dt=\partial/\partial t+V_{0x}\partial/\partial x\simeq
\partial/\partial t$ since the inflow velocity $V_{0x}$
is set to zero in our approximation.  

For the heating and cooling rates, we make use of ``realistic'' expressions 
that represent the physics of viscous accretion flows. Thus we write 
\beq 
Q^+=\alpha\frac{\rho c_s^2}{\Omega_K} \left(\frac{d
V_{0y}}{dx}\right)^2=4\alpha A^2\frac{\rho c_s^2}{\Omega_K}, \qquad
Q^-=-{\cal C}\rho^2\left(c_s^2\right)^n,
\label{heating}
\eeq 
where $\alpha\sim0.1$ is the standard Shakura-Sunyaev viscosity parameter,
$V_{0y}$ is the $y$-component of the unperturbed velocity, and ${\cal C}$ 
is a constant.  We leave the index $n$ in the cooling function
unspecified for now, but we note that $n=1/2$ corresponds to
non-relativistic free-free (bremsstrahlung) cooling.  

We assume that the perturbations in equations (\ref{eqs}) are of the
form $\exp(-i\omega t+ikx)$.  Substituting in the above equations and
solving, we obtain the following dispersion relation \citep{MN02}:
\bea 
\lefteqn{
\omega\left[\frac{\omega}{\gamma-1}+\frac{i(n-1)}{\tau_{\rm cool}}
+\frac{ik^2R^2}{\tau_{\rm cond}}\right] \left(\omega^2-\kappa_{\rm
epi}^2-k^2c_s^2\right){} } \nonumber\\ & &{}\qquad\qquad\quad
-\omega\left[\omega+\frac{i(2B/A-1)}{\tau_{\rm cool}}\right]k^2c_s^2=0,
\label{disp0}\label{disp1}
\eea
where 
\beq 
\tau_{\rm cool}=\left(\frac{\rho_0c_s^2}{Q_0^\mp}\right)
=\frac{\Omega_K}{4A^2\alpha}=\frac{4}{9\alpha s^2}\,\Omega_K^{-1}
\label{tau-cool}
\eeq 
is the cooling (heating) time of the gas and
\beq
\tau_{\rm cond}=\Omega_KR^2/\alpha_c c_s^2
\label{tau-cond}
\eeq is the conductive time scale. 

The dispersion relation (\ref{disp0}) corresponds to purely radial
perturbations.  The same relation can be used also for perturbations
in the vertical direction, except that we must set $\kappa_{\rm
epi}=0$.  Perturbations in the azimuthal direction are more
complicated.  Because of the shear, a non-axisymmetric wave packet is
distorted as a function of time, and must be analyzed by special
techniques which are beyond the scope of this study (see, e.g.,
\citealp{T77,GT78}).

Equation (\ref{disp0}) is a fourth-order polynomial and has four roots
corresponding to four modes.  A flow is unstable if any of the four
modes grows with time, i.e. if the corresponding root has 
${\rm Im}\;\omega>0$.  One of the roots of the dispersion relation is always
$\omega=0$.  This root corresponds to the viscous mode, which in the
present case is particularly simple because we neglected viscosity in
the momentum equation.  It is easy to show that if we introduce
viscosity into the momentum equation the viscous mode would become
stable, i.e., we will obtain ${\rm Im}\;\omega<0$.  We do not consider
the viscous mode further.

Let us now neglect thermal conduction for a moment, $\tau_{\rm cond}\to\infty$.
Then the physics of the remaining three modes may be understood by
considering equation (\ref{disp0}) in various limits.  Consider first
the limit $k\to0$.  In this limit, two of the roots are given by
$\omega=\pm\kappa_{\rm epi}$, corresponding to simple epicyclic
oscillations.  In the opposite limit $k\to\infty$, the same roots are
given by $\omega=\pm \gamma^{1/2}c_sk$, which shows that they
correspond to sound waves.  In the absence of heating and cooling
(i.e. $\tau_{\rm cool}\to\infty$), we can obtain an exact solution for
these roots which is valid for all $k$: 
\beq 
\omega^2=\kappa_{\rm epi}^2+{\gamma}c_s^2k^2.
\label{dispacoust} 
\eeq 
This is the standard
dispersion relation for sound waves in a differentially rotating flow.
The presence of $\gamma$ is because the relevant sound speed is the
adiabatic sound speed, $\gamma^{1/2}c_s$ (recall that $c_s$ is defined
to be the isothermal sound speed).

The final root of the dispersion relation (\ref{disp0}) corresponds to
the thermal mode.  In the limit $k\to0$, we obtain 
\beq
\omega=i\,(\gamma-1)\,\frac{(1-n)}{\tau_{\rm cool}}.  
\eeq 
We see that
the mode is stable (for $\gamma>1$) if $n>1$ and unstable if $n<1$.
In the opposite limit $k\to\infty$, we find 
\beq
\omega=i\,\frac{(\gamma-1)}{\gamma}\,\frac{(2-n-2B/A)}{\tau_{\rm cool}}.
\label{omega0}
\eeq
Therefore, the mode is stable if $n>2(1-B/A)$, i.e. $n>8/3$ for our
problem, and unstable otherwise.  Note that an accretion flow that is
cooled by free-free emission ($n=1/2$) is unstable in both limits.

\begin{figure}
\psfig{file=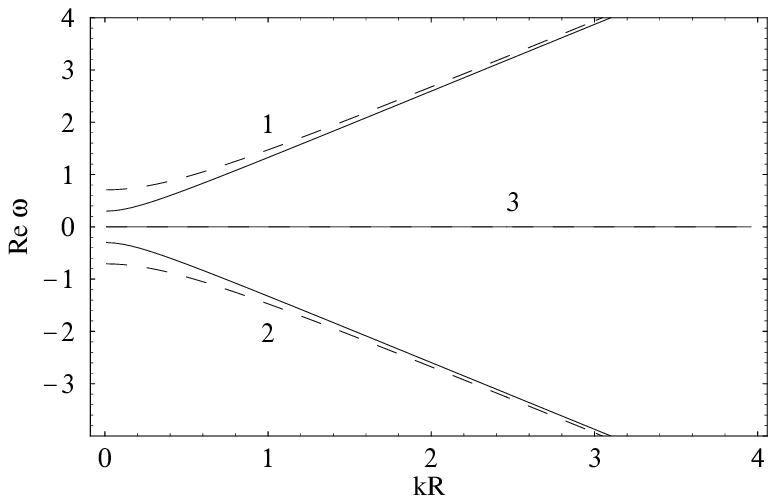,width=4in}\psfig{file=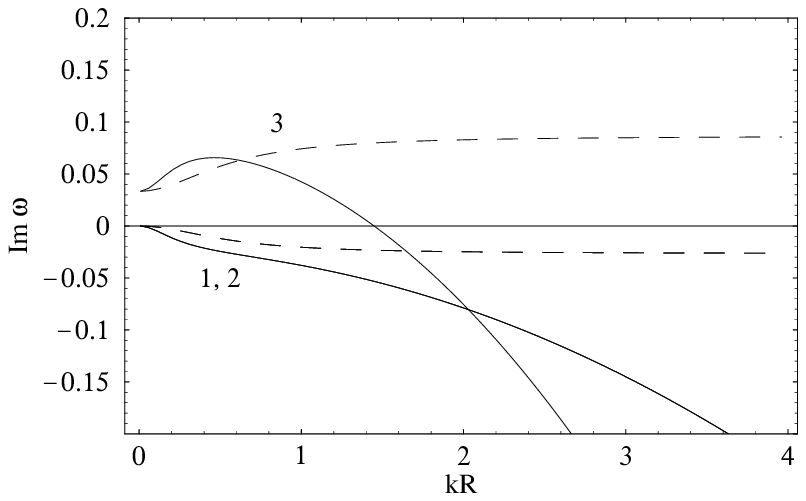,width=4in}
\caption{Real part (left panel) and imaginary part
(right panel) of the frequencies of three modes.  The curves labeled 1
and 2 refer to the two acoustic modes, and the curves labeled 3 refer
to the thermal mode.  The dashed curves correspond to the dispersion
relation (\ref{disp0}), which does not include thermal conduction
$\tau_{\rm cond}\to\infty$.
The following parameter values were used: $\kappa_{\rm
epi}/\Omega_K=s= \sqrt{0.5}$, $\tau_{\rm cool,ff}=10\Omega_K^{-1}$,
$c_s=\Omega_K R$, $\gamma=5/3$.  The mode frequencies are normalized
by the Keplerian frequency.  The solid curves correspond to the
dispersion relation (\ref{disp1}), which includes thermal conduction.
Here, $\tau_{\rm cond}=\tau_{\rm cool,ff}$, and the other parameters
are the same as before.}
\label{f:disp} 
\end{figure}

Let us now turn the conduction back on.
Figure \ref{f:disp} shows that the imaginary parts of all three modes
decrease rapidly with increasing $k$.  For the particular parameters
we have selected, the growth rate of the unstable thermal mode (curve
3) goes to zero at $k_{\rm crit}R\sim1.5$, and the mode is stable for
all $k>k_{\rm crit}$.

We may also analyze equation (\ref{disp1}) analytically. In the
large-$k$ limit, the root corresponding to the thermal mode is equal
to\footnote{ In deriving this equation from (\ref{disp1}) we have used
the fact that the acoustic time-scale is, in general, shorter than the
time-scale of the thermal mode, i.e., $\omega\ll k c_s$, and we have
neglected $\kappa_{\rm epi}$ as before. In this case we can neglect
$\omega^2$ in the second brackets, so that equation (\ref{omega1})
readily follows.  It may seem that this procedure fails when the
acoustic and thermal time-scales are comparable. This may happen when
$kR\sim1$ (for such perturbations, the sound crossing time is of order
the dynamical time) and when $\tau_{\rm cool}$ and $\tau_{\rm cond}$
are also comparable to the dynamical time, which is
$\sim\Omega_K^{-1}$. Nevertheless, even in this case, equation
(\ref{omega1}) works fairly well near the stability threshold. Indeed,
at the threshold itself, ${\rm Im}\ \omega=0$. Since further the
thermal mode frequency has no real part, we have $\omega\sim0$ near
the threshold; so we may safely neglect $\omega^2$ compared to
$k^2c_s^2$ in the second brackets of equation (\ref{omega1}).}  
\beq
\omega= i\,\frac{(\gamma-1)}{\gamma}\left(\frac{(2-n-2B/A)}{\tau_{\rm cool}} 
-\frac{k^2R^2}{\tau_{\rm cond}}\right).
\label{omega1}
\eeq 
Clearly, for large $k$, conduction stabilizes the thermal mode,
for the reasons explained at the beginning of this section.  Using the
above relation, we can estimate the critical $k_{\rm crit}$ above
which all $k$ are stable: 
\beq 
k_{\rm crit}^2R^2
=\frac{\tau_{\rm cond}}{\tau_{\rm cool}}\left(2-n-2\,\frac{B}{A}\right)
=\frac{13}{6}\,\frac{\tau_{\rm cond}}{\tau_{\rm cool}},
\label{stab-T}
\eeq 
where we have substituted $n=1/2$ (free-free cooling) and
$B/A=-1/3$ (Keplerian scaling).

We should comment here that in the theory described above, the
conductivity $\kappa_B$ is a quantity averaged over many field
correlation lengths.  Therefore, the results of the stability analysis
are valid only for perturbations on scales much larger than $l_B$.
(This is somewhat inconsistent since we have assumed that $l_B=\xi
R\sim0.1R$.)  For small-scale perturbations with $k\gg (\xi R)^{-1}$,
the local magnetic field is nearly homogeneous. Therefore, thermal
conductivity is anisotropic; its perpendicular component is of order
of $\kappa_B\sim n k_B v_T l_B \vartheta$ (for more discussion,
see, \citealp{MN02}), while the conductivity along
the field lines is much larger:
$$ 
\frac{q_\|}{q_\bot}=\frac{C\rho c_s^3\textrm{ sgn}(\nabla T)}
{-\alpha_c(\rho c_s^2/\Omega_K)(dc_s^2/dx)} \sim \frac{C\rho
c_s^3}{\alpha_c\rho c_s^3(H/R)} \sim\frac{5}{\alpha_c}\gg1, 
$$ 
as follows from equations (\ref{q-cond}) and (\ref{q-sat}).  
Thermal instability along the magnetic field lines is then 
suppressed more strongly than in the analysis presented above.

Thermal conduction in the hot brake flow is enormous \citep{MN02}
due to the very high temperature and large mean free path of particles.
The estimated value of thermal conduction is 
\beq
\alpha_c\simeq\textrm{ few }\times10^{-2}\xi_{-1}\vartheta_{-1}
\eeq
 Is this level of thermal
conduction enough to stabilize the thermal mode?

The relevant stability criterion is given in equation (\ref{stab-T}).
However, before we apply this criterion, we need to allow for the fact
that thermal conduction in the
flow is so strong that it modifies even the equilibrium structure of
the flow.  In particular, the cooling time (\ref{tau-cool}) is
modified  and becomes \citep{MN02}
\beq 
\tau_{\rm cool} =\frac{4}{(9\alpha
s^2+2\alpha_c)}\,\Omega_K^{-1}
\simeq\frac{2}{\alpha_c}\,\Omega_K^{-1}, 
\eeq 
where we have assumed
that $\alpha_c\gg\alpha s^2$, which is reasonable for typical
parameters, e.g. $\alpha\sim0.1,\ s\sim0.1$, $\alpha_c\sim0.1$.  From
equation (\ref{tau-cond}) and using the self-similar solution from Section
\ref{1st}, we obtain the thermal conductive time\footnote{
Alternatively, we recall that in the hot settling flow $H/R\sim 1$
and $H/R\simeq c_s/v_{\rm ff}\simeq c_s/(\sqrt2 \Omega_K R)$. Then
$\tau_{\rm cond}$ readily follows from equation (\ref{tau-cond}).}
\beq 
\tau_{\rm cond}=\frac{3}{\alpha_c}\,\Omega_K^{-1}, 
\eeq 
Substituting $\tau_{\rm cool}$ and $\tau_{\rm cond}$ into the
stability criterion (\ref{stab-T}), we then find 
\beq 
k_{\rm crit}R=\left[\frac{26\alpha_c}{9(9\alpha s^2+2\alpha_c)}\right]^{1/2}
\simeq\sqrt{\frac{13}{9}}\simeq1.2,
\label{kR-stab-ss}
\eeq 
that is, thermal modes with $kR\ga 1$ are stable.  This suggests
that the hot brake flow with thermal conduction is stable to the
thermal instability.

Whether the mode $kR=1$ itself is stable or not cannot be reliably
determined from our local analysis.  A global stability analysis is
necessary to properly account for the effects of geometry and
curvature, but this is beyond the scope of the present analysis.

\subsection{Observable properties}
\subsubsection{Spin-Up/Spin-Down of the Neutron Star}

The rate of spin-up of the accreting NS is given by
\beq
\frac{d}{dt}\left(I_*\Omega\right)=\dot J-\dot M\Omega(R_*)R_*^2 
\simeq-43s^3\alpha^2\dot M_{\rm Edd}\Omega_K(R_*)R_*^2,
\label{spin-up}
\eeq
where $I_*$ is the moment of inertia of the NS.  We have made use of
the fact that $|\dot J|\gg|\dot M\Omega(R_*)R_*^2|$ for the 
self-similar solution, and used equation (\ref{jdot}) for $\dot J$.
The negative sign in the final expression implies that the accretion
flow spins down the star. The above equation is for an unmagnetized NS. 
If the NS has a magnetosphere, the inner edge of the accretion flow is 
at the magnetospheric radius, $R_m$. In this case, let us define $s$ by 
$\Omega_*=s\Omega_K(R_m)=s\Omega_K(R_*)(R_m/R_*)^{-3/2}$. 
Substituting this in equation (\ref{spin-up}) with $I_*=constant$ 
and integrating, we obtain 
\beq
s=\frac{s_0}{\sqrt{1+t/\tau}}, \qquad
\tau=\frac{I_*}{86s_0^2\alpha^2\dot M_{\rm Edd}R_*^2}
\left(\frac{R_m}{R_*}\right)^{-3/2},
\eeq
where ${s_0=s({t=0})}$. The same result is valid for an unmagnetized NS
by setting $R_m=R_*$.
The quantity $\tau$ is the characteristic spin-down time of the NS. 
For a spherical NS of constant density,
$I_*=2M_*R^2_*/5=(0.8\times10^{33}\textrm{ g})mR_*^2$.
Substituting this expression, we obtain the spin-down rate
$\dot P_*/P_*=\tau^{-1}$ with 
\beq
\tau\simeq6.7\times10^{12}s^{-2}\alpha^{-2}
\left(\frac{R_m}{R_*}\right)^{-3/2}\textrm{ s}
=2\times10^{8}s_{0.1}^{-2}\alpha_{0.1}^{-2}
\left(\frac{R_m}{R_*}\right)^{-3/2}\textrm{ yr} .
\eeq 
Note the remarkable fact that the spin-down time scale is independent
of the mass of the NS, and the mass accretion rate! For the magnetic
case, the rate depends on the radius ratio ${R_m}/{R_*}$.

It is customary to express the spin-down rate as ${\dot P_*}/{P_*^2}$.
Writing 
\beq
P_*=\frac{2\pi}{s\Omega_K(R_*)}\left(\frac{R_m}{R_*}\right)^{3/2}
\eeq
and $\Omega_K(R_*)\simeq10^4m_{1.4}^{-1}\textrm{ rad/s}$, where 
$R_*=3R_g$ and $m_{1.4}=M_*/(1.4M_{\rm Sun})$, we obtain
\beq
\frac{\dot P_*}{P_*^2}
\simeq2.4\times10^{-10}m_{1.4}^{-1}\alpha^2s^3\textrm{ s}^{-2}
=2.7\times10^{-12}m_{1.4}^{-1}\alpha_{0.3}^2s_{0.5}^3\textrm{ s}^{-2},
\eeq
where $s_{0.5}=s/0.5$.
This spin-down rate is in good agreement with observational data
on the spin-down of X-ray pulsars for which \citet{YWV97} invoked ADAFs: 
4U~1626-67 has $\dot P/P^2\approx8\times10^{-13}\textrm{ s}^{-2}$ 
and $P=7.7\textrm{ s}$; OAO~1657-415 has 
$\dot P/P^2\approx2\times10^{-12}\textrm{ s}^{-2}$ and $P=38\textrm{ s}$,
and GX~1+4 has $\dot P/P^2\approx3.7\times10^{-12}\textrm{ s}^{-2}$ and 
$P=122\textrm{ s}$. Since the spin-down rate is quite sensitive to 
$\alpha$ and $s$, the observed data in individual systems can be fitted
by small adjustment of these parameters.

\subsubsection{Luminosity}

In computing the luminosity of the accretion flow, we must allow
for the energy release in both the boundary layer and the self-similar 
settling zone. We calculate their luminosities separately.

Radiation from the self-similar hot brake flow may be calculated
following the methods described by \citet{PN95} for a thin disk.
This method assumes that the luminosity at a given radius is determined 
by the local viscous energy production. This is a legitimate 
approximation for the settling flow in which $q^-=q^+$.
Keeping only the dominant terms, we find
\beq
L_{SS}=\frac{GM_*\dot M}{R_{in}}\left(1-js\right)
+\dot M\int_{P_{\rm in}}^{P_{\rm out}}\frac{dP}{\rho},
\eeq
where $R_{in}=R_*+\Delta_{BL}$ is the inner radius of the self-similar
zone and $\Delta_{BL}\ll R_*$ is the thickness of the boundary layer.
Here the first term inside the parentheses represents the luminosity
associated with the potential energy of the infalling gas,  
$L_{\rm pot}$, the second term $-js$ is the luminosity associated with 
the rotational energy extracted from the star, $L_{\rm rot}$ 
(note, $j<0$ in the self-similar solution), and the final integral 
is the ``enthalpy correction'', $L_{\rm enth}$. Using the analytical 
solution (\ref{sss})--(\ref{norm}) and assuming ${p_{\rm out}}=0$ for 
simplicity, we obtain
\bea
L_{\rm pot}&=&\dot M_{\rm Edd}c^2\frac{\mdot}{2r_*},\\
L_{\rm rot}&=&43\dot M_{\rm Edd}c^2\frac{\alpha^2}{2r_*} s^4,\\
L_{\rm enth}&=&-\dot M_{\rm Edd}c^2\frac{\mdot}{2r_*} .
\eea
Note that the leading terms in $L_{\rm pot}$ and $L_{\rm enth}$ 
cancel each other exactly. The luminosity of the self-similar flow 
is thus
\beq
L_{SS}\simeq
6.2\times10^{34}mr_3^{-1}\mdot_{-2}s_{0.1}^2
+8.9\times10^{33}mr_3^{-1}\alpha_{0.1}^2s_{0.1}^4 
\textrm{ erg\,s}^{-1},
\label{L-settl}
\eeq 
where $\mdot_{-2}=\mdot/0.01$, $s_{0.1}=s/0.1$, $r_3=r_*/3$,and we have assumed $s\ll1$.  Note that luminosity not associated 
with dissipation of rotational energy, represented by the first term
in equation (\ref{L-settl}), is much less than the commonly assumed
$\sim GM_*\dot M/R_*$. This is because the negative enthalpy
term has a large magnitude, as a result of the fact that the settling
flow is akin to a pressure supported, quasi-stationary atmosphere.

The second term in equation (\ref{L-settl}) is the luminosity of the 
settling zone.  Since the self-similar solution for this zone is 
independent of $\dot m$, the luminosity too shows no $\dot m$ dependence.  
Indeed, the luminosity remains finite even as $\dot m\to0$.  How is this
possible, and where does the energy come from?  The answer is that the
luminosity of the settling zone is supplied by the central star.  As
the star spins down, it does work on the accretion flow and the
energy released comes out as bremsstrahlung radiation.

The boundary layer luminosity requires a different method of
calculation since viscous energy production is negligible in this
zone: $\Omega\simeq$constant, and so
$q^+\propto\left(d\Omega/dR\right)^2\simeq0$.  As the accreting gas
cools in the boundary layer, starting from a nearly virial temperature
$\sim10^{12}$~K on the outside down to the NS temperature $\sim10^7$~K
near the surface, the thermal energy in the gas is emitted as
radiation.  To estimate the luminosity, we use the energy balance
equation, which is the sum of equations
(\ref{energy-p}),(\ref{energy-e}): 
\beq 
-q^-=\frac{\rho v}{\gamma-1}\frac{dc_s^2}{dR}-vc_s^2\frac{d\rho}{dR}
=\frac{\gamma}{\gamma-1}\rho v\frac{dc_s^2}{dR}-v\frac{dP}{dR}.  
\eeq
We can neglect the $dp/dR$ term because the pressure $p$ is essentially
constant in the boundary layer. To obtain the luminosity we integrate over
the boundary layer 
\beq 
L_{BL}=\int q^- 4\pi R^2 dR
=-\int\frac{\gamma}{\gamma-1}4\pi R^2\rho v\frac{dc_s^2}{dR}\, dR
=\frac{\gamma}{\gamma-1}\dot M\left(\Delta c_s^2\right).  
\eeq 
Since $c_s^2$ starts from nearly virial value and reaches close to zero, 
$\Delta c_s^2\simeq GM_*/R_*$. More precisely, 
$\Delta c_s^2=c^2\Delta(\theta_p+\theta_e)\simeq c^2\theta_{p0}r_*^{-1}$.
Therefore, the boundary layer luminosity is 
\beq
L_{BL}=\frac{\gamma}{\gamma-1}\dot M_{\rm Edd}c^2\frac{\mdot}{6r_*}
\approx 1.7\times10^{36}m\mdot_{-2}r_3^{-1},
\label{L-bl}
\eeq
where we have assumed $\gamma=5/3$. The total luminosity of the system
is $L=L_{SS}+L_{BL}$.  Note that $L_{BL}$ has the same scaling as the
first term in $L_{SS}$ [Eq. (\ref{L-settl})] and dominates the latter unless
$s\rightarrow1$.

\subsubsection{Effect of Comptonization}
\label{comptoniz}

Using the self-similar solution (\ref{sss}),(\ref{norm}), 
we may readily estimate the electron scattering optical depth and 
the $y$-parameter.\footnote{Here we just estimate where the effect
of Comptonization becomes significant. For better analytical 
approximations see, for instance, \citet{DLC91,TL95}. (In the 
latter paper, the expression for $y$ is not given, but it can be 
inferred using equation [24]:
$y=\tau_{es}[(\alpha+3)\theta/(1+\theta)+4d_0^{1/\alpha}\theta^2]$.)
Comptonization of free-free radiation has also been considered
by \citet{T88}.}
The optical depth is
\beq
\tau_{\rm es}\simeq\rho\kappa_{\rm es}R
\simeq10^3\alpha s^2r^{-1} \sim\alpha_{0.1}s_{0.1}^2r^{-1} ,
\eeq
where $\kappa_{\rm es}=\sigma_T/m_p$ is the electron scattering opacity for
ionized hydrogen. Since $r\ge3$, we see that $\tau_{\rm es}\le1/3$ for
reasonable parameters and the radiation is optically 
thin to electron scattering. The $y$-parameter is
\beq
y=16\theta_e^2\tau_{\rm es}
\simeq2\times10^6\alpha s^2r^{-2}\sim2\times10^3\alpha_{0.1}s_{0.1}^2r^{-2}.
\eeq
The radius at which $y\sim1$ is
\beq
r_c\sim45\alpha_{0.1}^{1/2}s_{0.1} .
\eeq
Above this radius the inverse Compton scattering is small and the 
self-similar solution is valid. For $r<r_c$, however, Comptonization is 
important and the electron temperature profile will be modified from 
the self-similar form. Since the electron-proton collisions are relatively 
weak (the plasma is two-temperature), other quantities, e.g., the density, 
proton temperature, etc., are unaffected. Comptonization is 
unimportant for low-viscosity flows, $\alpha\la0.01$ around slowly 
rotating NSs, $s\la0.01$, because then $r_c<r_*$.

\subsubsection{Spectrum}

We now estimate the spectrum of radiation emitted from the hot brake 
accretion flow. Let us neglect inverse Compton scattering for the moment.
The relativistic bremsstrahlung emissivity is 
approximated as 
\beq
\epsilon_\nu\propto\rho^2\exp{-(h\nu/kT_e)}
\textrm{ erg cm}^{-3}\textrm{ s}^{-1}\textrm{ Hz}^{-1}.
\eeq 
Therefore the luminosity per unit frequency is
\beq
L_\nu\propto\int_{R_*}^\infty\rho^2e^{-h\nu/kT_e}2\pi R^2\,dR 
\propto\int_{1/\nu_m}^\infty t^{-3}e^{-\nu t}dt
\propto\nu^2\Gamma(-2,\nu/\nu_m),
\eeq
where $\Gamma(a,z)=\int_{z}^\infty t^{a-1}e^{-t}dt$ is the incomplete
gamma-function and $\nu_m=kT_e(R_*)/h$ is the maximum frequency.
Above $\nu_m$ the spectrum falls exponentially and below $\nu_m$
it is nearly flat. We may, thus, replace the exponential in the integral with
a square function which is equal to unity for $\nu<\nu_m$ and 0 for 
$\nu>\nu_m$. With this approximation
\beq
L_\nu\simeq\frac{3}{2}\frac{L_{SS}}{\nu_m}\left(1-\frac{\nu^2}{\nu_m^2}\right),
\eeq
where $L_{SS}=\int L_\nu d\nu$ is the total luminosity of the self-similar 
flow, represented by equation (\ref{L-settl}). The break frequency,
$\nu_m$, is roughly given by  $h\nu_m\sim2.7\textrm{ MeV}$ for a typical 
electron temperature $T_{e,{\rm max}}\sim10^{10.5}~^\circ\textrm{K}$ 
[cf., equation (\ref{norm})]. 
At a typical x-ray energy, $h\nu\sim3\textrm{ keV}$, the observed
luminosity per decade is
\beq
\nu L_\nu\simeq1.7\times10^{31}m\alpha_{0.1}^2s_{0.1}^4
\left(\frac{h\nu}{3\textrm{ keV}}\right)\ \textrm{ erg s}^{-1},
\label{ss-spec}
\eeq
i.e., $\nu L_\nu\sim1.5\times10^{32}$ for a 300~Hz neutron star 
($s_{0.1}\sim1.6$). 
The spectrum is very hard with a photon index of order unity.  Therefore,
the luminosity per decade is much 
greater at higher photon energies and may be as high as 
$\sim\textrm{few}\times10^{34}-10^{35}\textrm{ erg/s}$ at 
$h\nu\sim\textrm{ MeV}$.

As shown in the previous section, Comptonization becomes important
below the radius $r_c$. At $r_c$, $y\approx1$ and the electron temperature is
\beq
T_e(r_c)\sim2.7\textrm{ MeV}/\sqrt{r_c}
\sim400\alpha_{0.1}^{-1/4}s_{0.1}^{-1/2}\textrm{ keV}.
\eeq
For $r<r_c$, the electron temperature will be determined 
self-consistently by Compton cooling rather than by bremsstrahlung
emission. Computing the spectrum from this region is beyond the
scope of the paper.  We also do not attempt to calculate the spectrum
of the radiation from the boundary layer.

\section{Conclusions}

In this paper we present the analytical self-similar solution 
describing this {\it hot brake flow}.
The hot brake flow exists at accretion rates as low as a few percent 
of Eddington or smaller. We studied its properties and showed that 
it is stable with respect to winds/outflows and convection, as well
as thermally stable (when the effects of thermal conduction are 
self-consistently included). The flow is subsonic everywhere.
The flow is cooling-dominated, it is powered by
the rotational energy of the central accretor which is braked by
viscous torques. A very interesting property of the flow is that,
except for the inflow velocity, all gas properties, such as density,
temperature, angular velocity, luminosity, and angular momentum flux,
are independent of the mass accretion rate (the flow properties do 
depend on the NS spin). This property implies that the density, 
temperature and angular velocity of the gas at any
radius are completely independent of the outer boundary conditions.
Therefore, such a flow cannot be matched to a general external medium.
Hence, there is a ``transition region'', represented by another
self-similar solution, which has an extra degree of freedom which 
permits the hot accretion flow to match general outer boundary conditions.
Matching the hot flow on the inside to general boundary conditions
at the star surface occurs via another transition region, known 
as the ``boundary layer''. A self-similar solution representing the 
boundary layer has also been derived.  

Although we presented the hot break solution in the context of a
neutron star, a similar accretion flow will certainly occur around 
a rapidly spinning white dwarf, and even around a spinning black hole, 
provided the \citet{BZ} mechanism is efficient enough
to power the MHD hot brake flow \citep{MMen02,Perna+03,MMur02}.

\end{document}